\documentclass[reprint, superscriptaddress]{revtex4-1}
\usepackage{amsmath, amssymb, amsfonts}
\usepackage{bm}
\usepackage{color}
\usepackage{float}
\usepackage{graphicx}% Include figure files
\usepackage{hyperref}% add hypertext capabilities
\usepackage{dcolumn}% align tables on decimal point

\hypersetup{
	colorlinks = false,
	urlcolor = {1 1 1}
}

\usepackage{color}

\begin{document}

\title{Benchmarking measures of network \\ controllability on canonical graph models}

\author{Elena Wu-Yan} 
\affiliation{Department of Computer and Information Sciences, University of Pennsylvania, Pennsylvania, PA 19104, USA}
\affiliation{Department of Bioengineering, University of Pennsylvania, Pennsylvania, PA 19104, USA}
\author{Richard F. Betzel}
\affiliation{Department of Bioengineering, University of Pennsylvania, Pennsylvania, PA 19104, USA}
\author{Evelyn Tang}
\affiliation{Department of Bioengineering, University of Pennsylvania, Pennsylvania, PA 19104, USA}
\author{Shi Gu}
\affiliation{Department of Bioengineering, University of Pennsylvania, Pennsylvania, PA 19104, USA}
\author{Fabio Pasqualetti}
\affiliation{Department of Mechanical Engineering, University of California, Riverside, CA, USA}
\author{Danielle S. Bassett}
\affiliation{Department of Bioengineering, University of Pennsylvania, Pennsylvania, PA 19104, USA}
\affiliation{Department of Electrical \& Systems Engineering, University of Pennsylvania, Pennsylvania, PA 19104, USA}
\email{dsb@seas.upenn.edu}

\date{\today}

\begin{abstract}

The control of networked dynamical systems opens the possibility for new discoveries and therapies in systems biology and neuroscience. Recent theoretical advances provide candidate mechanisms by which a system can be driven from one pre-specified state to another, and computational approaches provide tools to test those mechanisms in real-world systems. Despite already having been applied to study brain networks in several species, the practical performance of these tools and associated measures on simple networks with pre-specified structure has yet to be assessed. Here, we study the behavior of four control metrics (global, average, modal, and boundary controllability) on eight canonical graphs (including Erdos-Renyi, regular, small-world, random geometric, preferential attachment, and several modular networks) with different edge weighting schemes (Gaussian, power-law, and two nonparametric distributions from real-world brain networks). We observe that differences in global controllability across graph models are more salient when edge weight distributions are heavy-tailed as opposed to normal. In contrast, differences in average, modal, and boundary controllability across graph models (as well as across nodes in the graph) are more salient when edge weight distributions are less heavy-tailed. Across graph models and edge weighting schemes, average and modal controllability are negatively correlated with one another across nodes; yet, across graph instances, the relation between average and modal controllability can be positive, negative, or non-significant. Collectively, these findings demonstrate that controllability statistics (and their relations) differ across graphs with different topologies, and that these differences can be muted or accentuated by differences in the edge weight distributions. More generally, our numerical studies motivate future analytical efforts to better understand the mathematical underpinnings of the relationship between graph topology and control, as well as efforts to design networks with specific control profiles.

\end{abstract}

\maketitle

\textbf{Many real-world systems are composed of many individual components that interact with one another in a complex pattern to produce diverse behaviors. Understanding how to intervene in these systems to guide behaviors is critically important to facilitate new discoveries and therapies in systems biology and neuroscience. A promising approach to optimizing interventions in complex systems is network control theory, an emerging conceptual framework and associated mathematics to understand how targeted input to nodes in a network system can predictably alter system dynamics. While network control theory is currently being applied to real-world data, the practical performance of these measures on simple networks with pre-specified structure is not well understood. In this study, we benchmark measures of network controllability on canonical graph models, providing an intuition for how control strategy, graph topology, and edge weight distribution mutually depend on one another. Our numerical studies motivate future analytical efforts to gain a mechanistic understanding of the relationship between graph topology and control, as well as efforts to design networks with specific control profiles.}

\section*{Introduction}
Complex systems can be modeled as networks in which the system's elements and their pairwise interactions are represented, respectively, as nodes and edges in a graph \cite{newman2011structure}. Drawing on a subfield of mathematics known as graph theory, network analysis allows for the quantification of a system's topological organization and offers insight into its function. Network models and associated graph representations have been adopted in a range of disciplines to successfully investigate the structure and function of social, economic, and biological systems \cite{carrington2005models, schweitzer2009economic, goh2007human}.

Complex systems are also dynamic, meaning that their elements can be associated with internal states that evolve and fluctuate over time \cite{boccaletti2006complex, porter2014dynamical}. The state of an element is system-dependent and can correspond to any number of real-world observables, including disease status (e.g., ``healthy'' or ``infected'') \cite{newman2002spread, colizza2006role}, the concentration of nutrients at a particular site \cite{heaton2012analysis, papadopoulos2016embedding}, the electrical activity of neurons \cite{bettencourt2007functional,teller2014emergence,wiles2017autaptic} or hemodynamic activity of brain areas \cite{goni2014resting, honey2007network}. The evolution of a system's state over time depends upon the organization of its underlying network. For example, in social systems, individuals become exposed to and infected by disease through their social contacts \cite{ray2016network}. Similarly, in neural systems, activity propagates from neuron-to-neuron or region-to-region along axonal projections and fiber bundles \cite{muldoon2016stimulation}.

An important question (and one that has been the topic of many recent inquiries \cite{liu2011controllability, yan2012controlling, sun2013controllability, pasqualetti2014controllability, ruths2014control}) is whether a networked, dynamical system can be controlled \cite{liu2016control}. Network control refers to the prospect of selectively influencing the evolution of a system's state by introducing time-varying external input(s) that drive it from one state to another along a particular trajectory. Current efforts are actively tackling diverse challenges in developing a framework to determine whether control is theoretically possible for a given system \cite{liu2011controllability}, identifying nodes and edges that are important for efficient control \cite{kim2016topological}, and proposing realistic strategies for enacting control over complex systems \cite{cornelius2013realistic}.

\begin{figure*}
	\begin{center}
		\centerline{\includegraphics[width=0.6\textwidth]{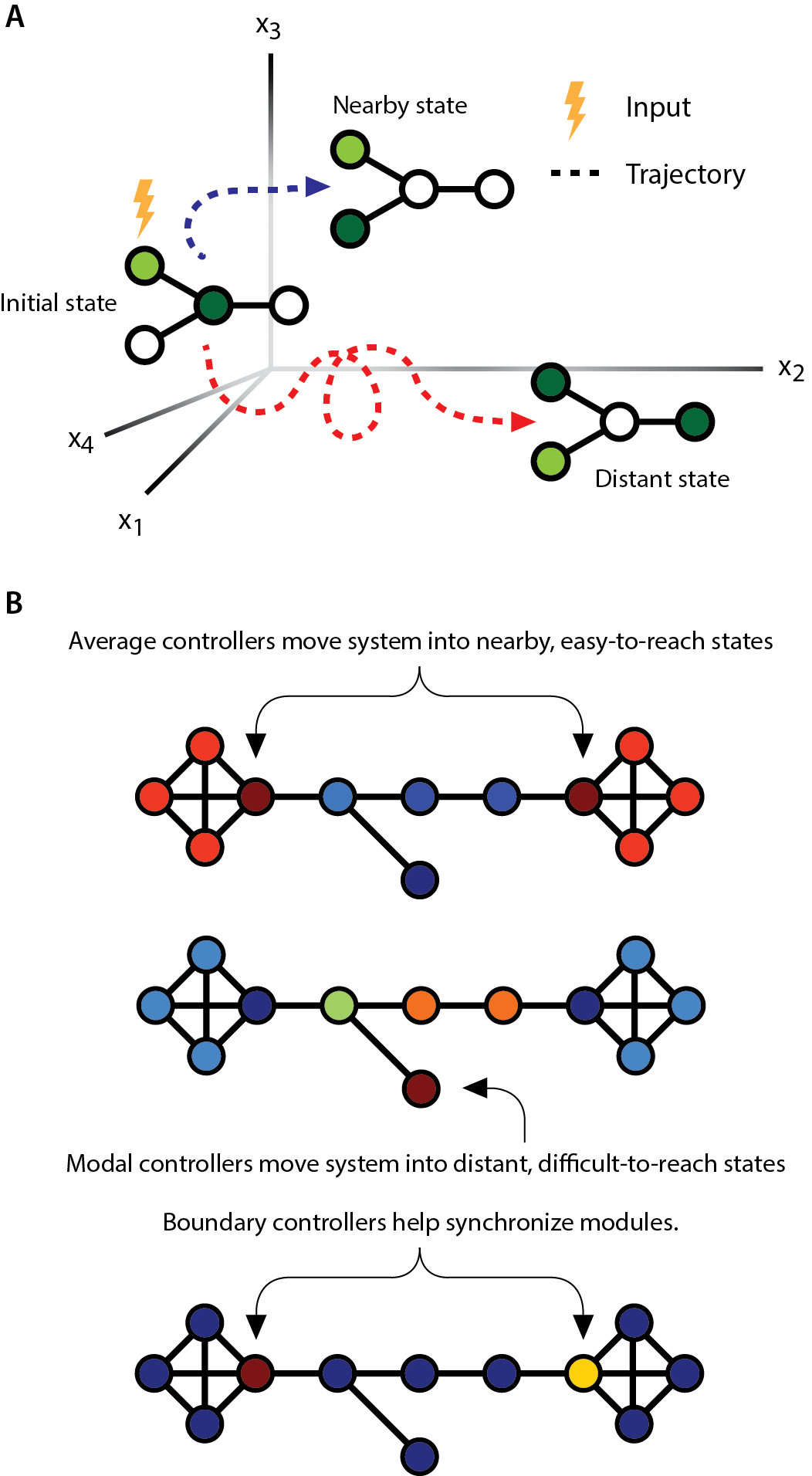}}
		\caption{\textbf{Network control theory and metrics to describe control strategies.} \emph{(A)} Controlling a network corresponds to taking a network from a given state (defined as a pattern of node activation magnitudes) to another state (a different pattern of node activation) by applying a control input to a particular node (or set of nodes) in the network. The set of states that is traversed from initial state to target state is known as the trajectory. The target state can be either nearby or distant in energy, and control input is placed at different nodes to effect these two different control strategies. \emph{(B)} Metrics to quantitatively characterize different control strategies include average, modal, and boundary controllability. \emph{(Top)} Average controllability identifies nodes that can move the system into nearby easy-to-reach states with little control energy.  \emph{(Middle)} Modal controllability identifies nodes that can move the system into distant, difficult-to-reach states. \emph{(Bottom)} Boundary controllability identifies nodes that can help to synchronize or desynchronize communities in the network. } \label{fig_schematic}
		\vspace*{0mm}
	\end{center}
\end{figure*}

One particularly important approach involves quantifying the contributions of control points (nodes) in driving a system through state space \cite{pasqualetti2014controllability}. Based on the topological organization of a network as measured by the configuration of edges, certain nodes may be predisposed to drive the system in particular manners based on \emph{control strategies} (see Fig.~\ref{fig_schematic}). For example, some nodes might be better at driving the system into a multitude of different, easy-to-reach states (\emph{average controllability}) while others may be well-suited for driving the system into difficult-to-reach states (\emph{modal controllability}). Still others are situated between different modules and, therefore, have the capacity to regulate and control inter-modular synchronization (\emph{boundary controllability}). These measures of controllability can be made on virtually any networked system, but have been applied most successfully to study large-scale human brain networks \cite{gu2015controllability, tang2016structural}. These early studies demonstrated that brain systems that support different types of function are also characterized by unique controllability profiles, and that these profiles follow distinct trajectories across late development. Despite their application to the study of real-world brain networks, the behavior of average, modal, and boundary controllability measures in practical contexts on canonical network models has not been explored. Such an investigation would help contextualize the behavior of these measures on real-world networks.

In this report, we study the behavior of four metrics that characterize overall controllability (global) as well as distinct control strategies (average, modal, and boundary controllability) in common graph models. We choose these models both to canvas architectures that are  proven  benchmarks in the analysis of complex systems generally, and also to probe architectures that are particularly relevant for neural networks in human and non-human species, a question of critical importance in recent studies \cite{klimm2014resolving,samu2014influence,roberts2016contribution,henderson2011geometric,betzel2017modular}. This rationale motivated our study of graph ensembles drawn from random, regular lattice, small-world, random geometric, and several modular graph models with differing module size. Because many real-world networks are better characterized by weighted graphs than binary graphs, including brain networks \cite{bassett2016small}, we assigned weights to the edges in each of these models by drawing from theoretically defined distributions such as Gaussian and power-law, as well as from empirically measured distributions including the fractional anisotropy along white matter tracts in the human brain, or the number of streamlines tracked between brain regions. Together, these choices provide a multidimensional space in which to gain understanding for how control strategy, graph topology, and edge weight distribution mutually depend on one another, thereby informing future efforts to design networks with specific control profiles.

\section*{Materials and Methods}

\subsection*{Network generation.}

\subsubsection*{Network definition.}

Following common parlance, we will use the terms \emph{graph} and \emph{network} interchangeably to refer to a set of nodes connected by edges. Mathematically, we represent a graph $G$ as $G = (V, E)$, where $V$ is the set of vertices, or nodes, and $E$ is the set of edges between nodes. Computationally, we represent graphs as adjacency matrices. A matrix with $\vert V \vert = N$, or $N$ nodes, has an adjacency matrix $A_{ij}$ of size $N \times N$, where each element $A_{ij}$ denotes the connection strength, or edge weight, between node $i$ and node $j$. We study undirected graphs represented by symmetric adjacency matrices, where $A_{ij} = A_{ji}$, with no self loops -- connections of a given node with itself -- i.e. for all $i$, $A_{ii} = 0$. Finally, we distinguish between (i) a binary graph, which only has edge weight values of 0 or 1 to denote the absence or presence, respectively, of a connection between nodes, and (ii) a weighted graph, which can have scaled edge weight values denoting the strength of node-node relations.

\subsubsection*{Network models}

Historically, the fields of graph theory and network science have found it useful to define specific models of graph architectures based on rules for wiring, rules for growth, rules for pruning, and rules for explaining patterns of missing data. In each case, such models provide the grounds for developing and benchmarking novel network statistics. Here, we take a few of the most common graph models from the general literature, as well as more specifically from the literature postulating models of the topology observed in human brain networks \cite{klimm2014resolving,samu2014influence,roberts2016contribution,henderson2011geometric}, and use them to benchmark the behaviors of recently defined measures of network controllability \cite{pasqualetti2014controllability}. Below, we briefly describe the 8 graph models that we chose as the basis for our analysis of controllability.

\begin{itemize}
	\item \textit{(WRG) Weighted Random Graph model}: Arguably the most fundamental, this graph model is a weighted version of the canonical Erdős–Rényi model. For all pairs of $N$ nodes, we modeled the weight of the edge by a geometric distribution with probability of success $p$, the desired edge density of the graph. Each edge weight was assigned the number of successes before the first failure \cite{sizemore2017classification}.
	\item \textit{(RL) Ring Lattice model}: In contrast to the random nature of the WRG, the ring lattice model is one with strict order. We arranged $N$ nodes on the perimeter of a regular polygon, each with degree $k$, determined by the desired edge density. We then connected each node to the $\frac{k}{2}$ nodes directly before and after it in the sequence of nodes on the polygon. Edge weights were assigned the inverse of the path length between the two nodes, determined by traversing the perimeter of the polygon \cite{sizemore2017classification}.
	\item \textit{(WS) Watts-Strogatz model}: A model that bridges both the order of the RL and the disorder of the WRG, the Watts-Strogatz graph model is a ring lattice model in which edges are rewired uniformly at random to create a small-world network. Following \cite{sizemore2017classification}, we chose the probability $r$ of rewiring a given edge to maximize the small-world propensity \cite{muldoon2016small}.
	\item \textit{(MD2) Modular Network with 2 communities model}: While the previous models can display some local clustering structure, they lack meso-scale organization in the form of modules or communities. In contrast, the Modular Network with 2 communities model is a graph of $N$ nodes and $K$ edges placed so as to form 2 communities. Pairs of nodes \textit{within} communities are connected with edge density 0.8, and pairs of nodes \textit{between} communities (where one node in the pair is in one community and the other node in the pair is in a different community) are connected to fulfill the desired total edge density $p$. We assigned weights to existing edges by considering a geometric distribution with probability of success $p$ if the nodes were in the same module and $1-p$ if the nodes were in different modules. Each edge weight was assigned the number of successes before the first failure \cite{sizemore2017classification}.
	\item \textit{(MD4) Modular Network with 4 communities model}: This model is generated in a manner identical to that used in the MD2 graph model, with the exception that MD4 has 4 communities.
	\item \textit{(MD8) Modular Network with 8 communities model}: This model is generated in a manner identical to that used in the MD2 graph model, with the exception that MD4 has 8 communities.
	\item \textit{(RG) Random Geometric model}: In contrast to most of the previous graph models that were agnostic to any embedding space, the Random Geometric model contains $N$ nodes, chosen randomly from a unit cube, and edges whose weights were equal to the inverse of the Euclidian distance between two nodes. We kept only the $K$ shortest edges, in order to maintain the desired edge density $p$  \cite{sizemore2017classification}.
	\item \textit{(BA) Barabási–Albert model}: In our final graph model, we use software from \cite{klimm2014resolving} to generate a typical BA model -- a scale-free network that exhibits preferential attachment to existing nodes of high degree -- with $N$ nodes and $K$ edges. Each edge weight was assigned the average degree of the two nodes it connected.
\end{itemize}

While of course this is not an exhaustive list of the possible graph models that one might wish to study, we focus on this set because it provides a useful assessment of quite different topologies, and because most of these models have been suggested as relevant benchmarks against which to compare brain networks in previous studies.

\subsubsection*{Network size.}

In all of the graph models described above, two parameters must be fixed \emph{a priori}: the number of nodes $N$ in the network, and the number of edges $K$ in the network. We chose the number of nodes to be 128 (see Results section below), and we confirmed consistency of our findings across these other network sizes (256, or 512, see Supplementary Materials). We chose the number of edges to produce network densities that were consistent with those observed empirically in large-scale human brain graphs. Specifically, drawing on recently published data from 30 healthy adult individuals by capitalizing on a 19-minute multiband diffusion spectrum imaging sequence \cite{betzel2016optimally}, we assigned the 128-node graphs an edge density of 0.2919; the 256-node graphs an edge density of 0.2175; and the 512-node graphs an edge density of 0.1396. For each network size, we generated 100 instantiations of each of the 8 graph models described above.

\subsubsection*{Network weighting.}

All of the 8 graph models described above were weighted graph models \cite{sizemore2017classification}. While it is important to study weighted (as opposed to binary) graph models to benchmark network controllability statistics that are currently being applied to real-world weighted graphs, comparisons across models are confounded by the fact that each model can have a very different edge weight distribution. Here, we sought to disentangle the impact of graph model from the impact of edge weight distribution on network controllability statistics. Practically, we therefore developed a pipeline to reweight all of the graph models fairly, and with a fixed edge weight distribution.

We began by adding random noise on the order of $10^{-7}$ to all edge weights in all network models; this process ensures the uniqueness of each edge weight, while maintaining the relative weight magnitudes. Next, we sorted edges by weight, and then replaced each edge with corresponding ordered values pulled from a specific edge weight distribution of interest, of which we defined four. The first was a Gaussian distribution with a mean of 0.5 and a standard deviation of 0.12. The second was a power law distribution with a slope of -3 and a range of values from $10^{-5}$ to $10^{5}$. Both Gaussian and power law distributions are ubiquitously found in real-world networks, and in fact form natural benchmarks for edge weight distributions taken from neuroimaging data. The third and fourth edge weight distributions of interest were taken from \cite{betzel2016optimally} to closely model empirical weighting distributions in large-scale human brain structural networks estimated from diffusion imaging tractography. Specifically, these two distributions were streamline counts (normalized by the geometric mean of regional volumes) and fractional anistropy (FA). Importantly, the reweighting scheme we describe here allowed us to use the same edge weighting across all graphs to guarantee that differences in controllability were due to topology and not to other properties of the graphs, like differing edge weights and scaling.

\subsubsection*{Network ensembles.}

In summary, we study three network sizes (128, 256, or 512 nodes) for each of the four edge weightings (Streamline Counts, FA, Gaussian, Power Law), thus totaling 12 sets of networks, each of which included 100 instantiations of each of the 8 graph models. We next turn to an examination of network controllability statistics in these 12 sets of weighted graphs.

\subsection*{Network Controllability.}

\subsubsection*{A linear model of brain dynamics.}

Although the relationship of brain structural networks to the correlation structure of spontaneous neural activity is, in general, non-linear, a great deal of variance in that correlation structure can, nonetheless, be explained by simple linear models \cite{galan2008network}. Accordingly, we define brain dynamics with a noise-free, linear, discrete-time, and time-invariant network model:

\begin{equation}
\mathbf{x}(t + 1) = \mathbf{A}\mathbf{x}(t) + \mathbf{B}_{\mathcal{K}}\mathbf{u}_{\mathcal{K}}(t),
\label{eq1}
\end{equation}

\noindent where $\mathbf{x} : \mathbb{R}_{\ge 0} \rightarrow \mathbb{R}^N$ denotes a brain region's state (i.e., its magnitude of electrical or hemodynamic activity) and $\mathbf{A}$ is the symmetric, weighted adjacency matrix. The input matrix, $\mathbf{B}_{\mathcal{K}}$, identifies control points, $\mathcal{K} = \{ k_1 , \ldots , k_m \}$, is defined as:

\begin{equation}
\mathbf{B}_\mathcal{K} = [e_{k_1} , \ldots , e_{k_m}]
\end{equation}

\noindent where $e_i$ indicates the $i$-th canonical vector (of length $N$). The input, $\mathbf{u}_\mathcal{K} : \mathbb{R}_{\ge 0} \rightarrow \mathbb{R}^m$, specifies the strategy for controlling these dynamics. Note that in Eq.~\ref{eq1}, we scaled the elements of $\mathbf{A}$ by $1 / (1 + \lambda_{\text{max}})$ (where $\lambda_{\text{max}}$ is the largest eigenvalue of unscaled $\mathbf{A}$), which ensures that the scaled version of $\mathbf{A}$ is Schur stable (i.e., all eigenvalues of $\mathbf{A}$ are $<1$ in magnitude). We note that different choices for the normalization will accentuate \emph{versus} de-emphasize different scales of dynamics, and it will be interesting in future to study how the choice of normalization impacts observed patterns of controllability.

\subsubsection*{Global controllability.}

Under these dynamics, we define a series of control metrics that quantify different intuitions of controllability. The first is a measure of \emph{global controllability}, which is used to assess whether control (i.e., the ability to steer the system from any arbitrary network state to any other arbitrary network state), is theoretically possible. In general, the ease or difficulty of control is related to the structure and eigenvalues of the controllability Gramian:

\begin{equation}
\mathbf{W}_\mathcal{K} = \sum_{t = 0}^\infty \mathbf{A}^t \mathbf{B}_\mathcal{K} \mathbf{B}_\mathcal{K}^T \mathbf{A}^t,
\end{equation}
where the eigenvalues of the inverse Gramian indicate ease of control. \noindent We define \emph{global controllability} to be the smallest eigenvalue of $\mathbf{W}_\mathcal{K}$. The structure of $\mathbf{W}_\mathcal{K}$ depends upon both the topological organization of a network (which we assume to be fixed) and the set of nodes assumed to be control points, i.e. $\mathbf{B}_\mathcal{K}$. As in \cite{gu2015controllability}, we test all possible single-point control strategies (where $\mathbf{B}_\mathcal{K}$ is a one-dimensional column vector), computing $\mathbf{W}_\mathcal{K}$ and its corresponding eigenvalues along with the global controllability metric for each case. Accordingly, for every realization of any graph model, we obtain $N$ global controllability scores for each of the $N$ nodes.

\subsubsection*{Average and modal controllability.}

In addition to global controllability, we also characterized networks using three other node-level controllability metrics. The first, \emph{average controllability}, describes the ease with which control points in a network can move the system into nearby, easily reached states, and is computed as $trace(\mathbf{W}_\mathcal{K}^{-1})$. The second metric, \emph{modal controllability}, quantifies the ease with which control points in a network can move the system into distant, hard-to-reach states. Let $\mathbf{V} = [v_{ij}]$ be the eigenvector matrix of $\mathbf{A}$. Then the modal controllability of node $i$ is defined as:

\begin{equation}
\phi_i = \sum_{j = 1}^N (1 - \lambda_j^2(\mathbf{A})) v_{ij}^2
\end{equation}

\noindent where $\lambda_j(\mathbf{A})$ are the eigenvalues of the scaled version of $\mathbf{A}$.

Past studies have shown that, for a given graph, average and modal controllability tend to be anti-correlated with one another across nodes \cite{gu2015controllability, tang2016structural}, so that brain regions well-suited for moving the system into easy-to-reach states are also poorly-suited for moving the system into difficult-to-reach states and \emph{vice versa}.

\subsubsection*{Boundary controllability.}

Lastly, we also investigated \emph{boundary controllability}, which describes the ease with which control points in a network can act to decouple or connect communities (subnetworks) within the system. To detect boundary controllers, we followed the procedure described in \cite{gu2015controllability}, which modified the algorithm proposed in \cite{pasqualetti2014controllability}. Briefly, this procedure involves iteratively bi-partitioning the original network into progressively smaller subnetworks and, at each level, identifying nodes whose connections span both halves. In \cite{pasqualetti2014controllability}, the initial partition of the network into two subnetworks was generated based on the Fiedler eigenvector of $\mathbf{A}$. Here, because brain networks are composed of many distinct communities \cite{sporns2016modular}, we initialize the algorithm instead with a partition obtained by maximizing a modularity quality function \cite{newman2004finding}:

\begin{equation}
Q = \sum_{ij} [A_{ij} - P_{ij}] \delta( \sigma_i \sigma_j )
\end{equation}

\noindent where $P_{ij}$ is the expected weight of the connection between nodes $i$ and $j$, $\sigma_i$ is the subnetwork (community) assignment of node $i$, and $\delta( \sigma_i, \sigma_j )$ is the Kronecker delta function, which equals one when $\sigma_i = \sigma_j$ and zero otherwise. Here, the expected weight of connections is determined based on a null model in which nodes' strengths are preserved but connections are, otherwise, formed at random: $P_{ij} = \frac{k_i k_j}{2m}$, where $k_i = \sum_j A_{ij}$ and $2m = \sum_i k_i$. This null model is commonly known as the Newman-Girvan null model.

The subnetworks defined by this initial partition were then iteratively bi-partitioned according to the Fiedler eigenvector. With each division, boundary points were identified as nodes that maintained supra-threshold connections to both subnetworks. The value of this threshold, $\rho = 1 \times 10^{-5}$, was selected so as to -- in general across graph models and instantiations -- maximize the number of unique values of boundary controllability across nodes in a given graph. To demonstrate the reliability of our results given reasonable variations in this choice, we also provide results for  $\rho = 1 \times 10^{-3}$ and $\rho = 1 \times 10^{-8}$ in the Supplementary Materials.

\section*{Results}

\subsection*{Variation in network controllability statistics across graph models}

We begin by examining how network controllability statistics vary over graphs within a given ensemble, and whether that variation differs as a function of graph model. Following the procedure outlined in the Methods section, for each controllability type, we took the mean of the 128 sorted controllability values across the $n$ nodes in each graph instance, giving us 100 controllability values averaged over the 128 regions. For each controllability type, we used four identical edge weight distributions corresponding to fractional anisotropy (FA) and streamline counts (SC) from real brain data, a Gaussian distribution, and a power law distribution.

\begin{figure*}
	\begin{center}
		\centerline{\includegraphics[width=0.9\textwidth]{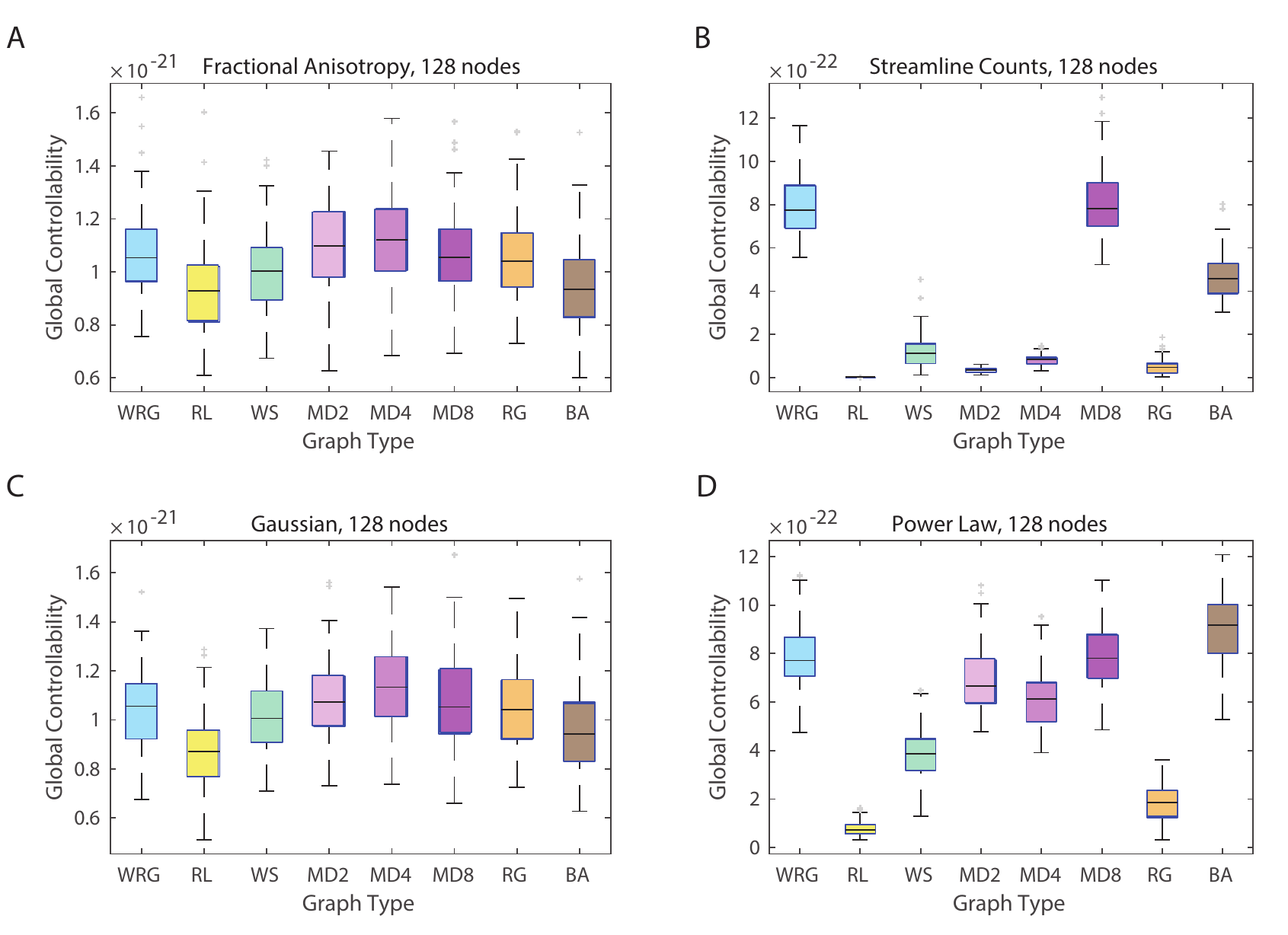}}
		\caption{\textbf{Variation in global controllability as a function of edge weighting and graph model.} Global controllability values were averaged over nodes in each graph model ensemble, and therefore boxplots show variation over graphs in the ensemble. Results are shown for four edge weighting schemes: \emph{(A)} FA, \emph{(B)} streamline counts, \emph{(C)} Gaussian, \emph{(D)} power-law. The eight graph models include the weighted random graph (WRG), the ring lattice (RL), the Watts-Strogatz small-world (WS), the modular graphs (MD2, MD4, MD8), the random geometric (RG), and the Barabasi-Albert preferential attachment (BA) models. } \label{fig_g128n}
		\vspace*{-10mm}
	\end{center}
\end{figure*}

When considering global controllability for graph models constructed with Gaussian and FA weighting schemes, we observed that all graph types had global controllability values similar in range and high variance, with RL at the low end of the range and MD4 at the high end of the range (Fig.~\ref{fig_g128n}A,C). For graph models constructed with steamline counts and power-law weighting schemes, we observed varying behavior. The SC-weighted RL, WS, MD2, MD4, and RG graphs all had low mean and low variance, while WRG, MD8, and BA had higher mean values and higher variance. The power law-weighted RL, WS, and RG graphs had lowest means and variance, while the other graph types had higher mean and variance, with BA having the highest mean and variance (Fig.~\ref{fig_g128n}B,D). One-way ANOVAs indicated significant effects of graph type for all of the four edge weight distributions (see Table~\ref{tab:anovas1}).

\begin{figure*}
	\begin{center}
		\centerline{\includegraphics[width=0.9\textwidth]{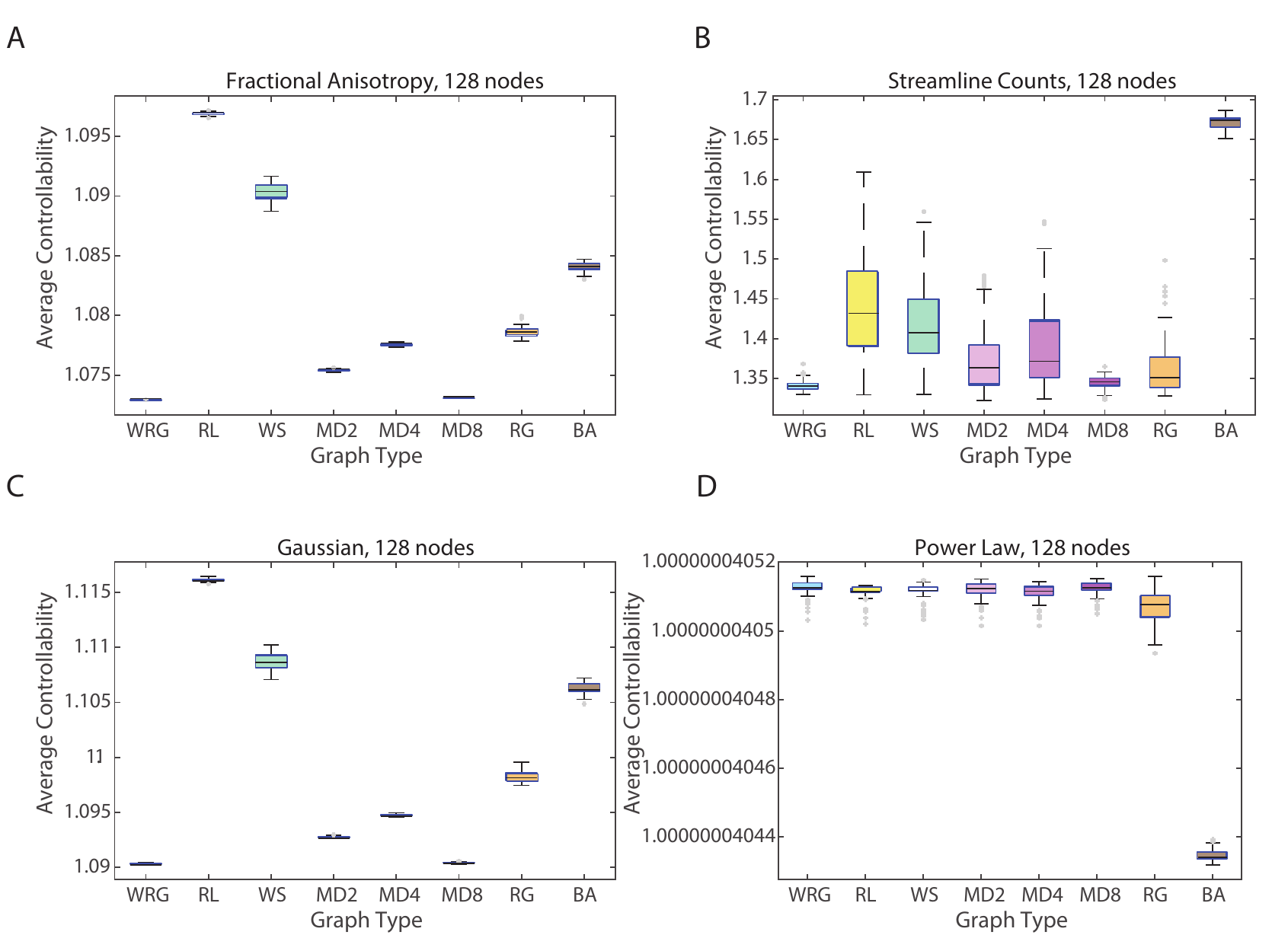}}
		\caption{\textbf{Variation in average controllability as a function of edge weighting and graph model.} Average controllability values were averaged over nodes in each graph model ensemble, and therefore boxplots show variation over graphs in the ensemble. Results are shown for four edge weighting schemes: \emph{(A)} FA, \emph{(B)} streamline counts, \emph{(C)} Gaussian, \emph{(D)} power-law. The eight graph models include the weighted random graph (WRG), the ring lattice (RL), the Watts-Strogatz small-world (WS), the modular graphs (MD2, MD4, MD8), the random geometric (RG), and the Barabasi-Albert preferential attachment (BA) models. } \label{fig_a128n}
		\vspace*{-10mm}
	\end{center}
\end{figure*}

When considering average controllability for graph models constructed with Gaussian and FA weighting schemes, we observed that WRG, MD2, MD4, and MD8 all had similarly low mean average controllability values (between 1.090 and 1.095) with small variance (Fig.~\ref{fig_a128n}A,C). RG, BA, WS, and RL then followed in increasing order of mean, and WS had the highest variance. For graph models constructed with the power-law weighting scheme, we observed relatively uniform behavior, all with high controllability values, low variance, and skewed left-tailed distributions toward lower controllability values (Fig.~\ref{fig_a128n}D). The exception was the BA graph, which had low average controllability and low variance. The SC-weighted graphs varied in mean value and variance, with WRG and MD8 having the lowest mean and lowest variance and BA having the highest mean and low variance (Fig.~\ref{fig_a128n}C). RG, MD2, MD4, WS, and RL then followed with increasing mean and variance, all with slightly right-tailed distributions. One-way ANOVAs indicated significant effects of graph type for all of the four edge weight distributions (see Table~\ref{tab:anovas1}).

\begin{figure*}
	\begin{center}
		\centerline{\includegraphics[width=0.9\textwidth]{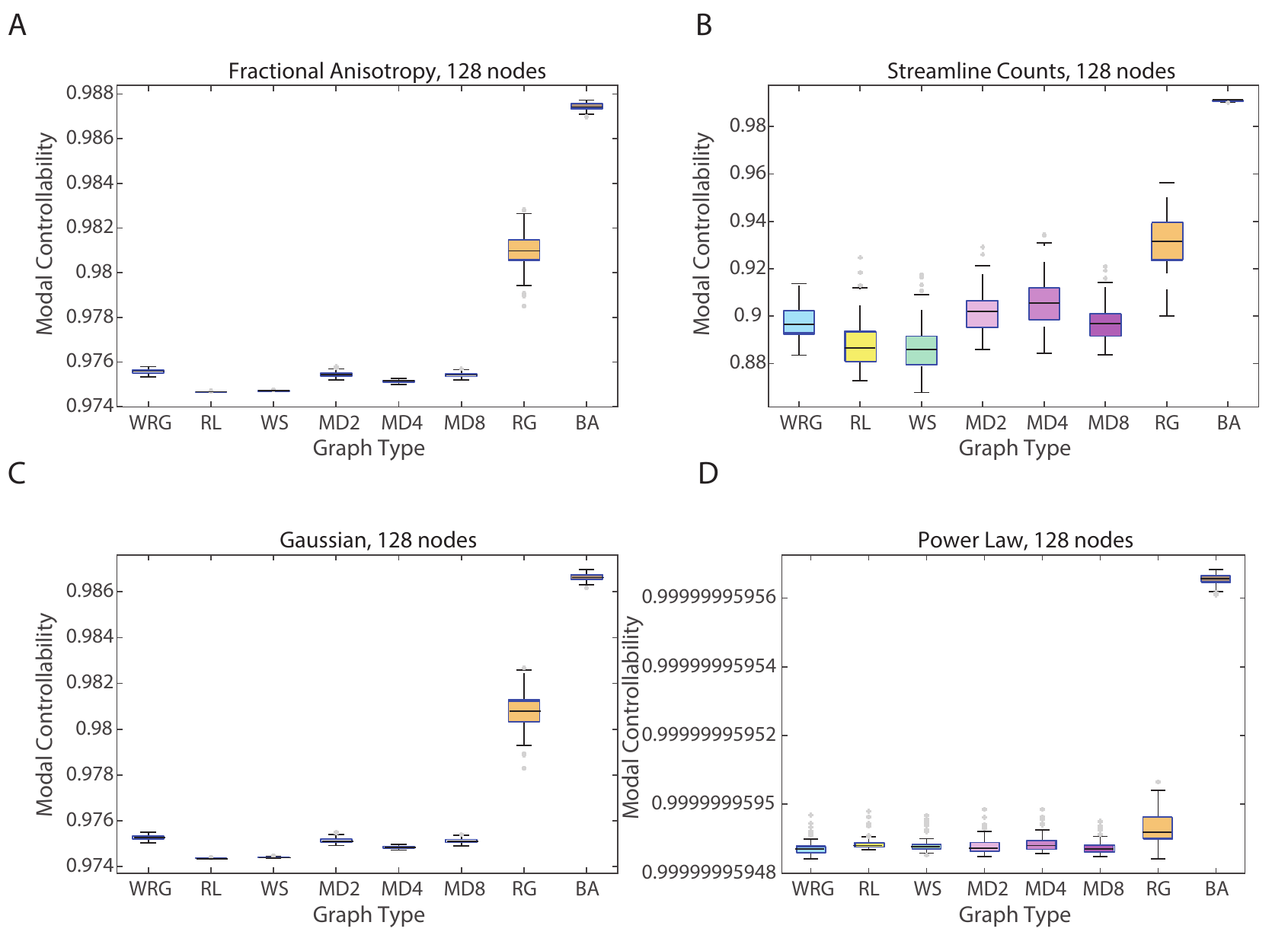}}
		\caption{\textbf{Variation in modal controllability as a function of edge weighting and graph model.} Modal controllability values were averaged over nodes in each graph model ensemble, and therefore boxplots show variation over graphs in the ensemble. Results are shown for four edge weighting schemes: \emph{(A)} FA, \emph{(B)} streamline counts, \emph{(C)} Gaussian, \emph{(D)} power-law. The eight graph models include the weighted random graph (WRG), the ring lattice (RL), the Watts-Strogatz small-world (WS), the modular graphs (MD2, MD4, MD8), the random geometric (RG), and the Barabasi-Albert preferential attachment (BA) models. } \label{fig_m128n}
		\vspace*{-10mm}
	\end{center}
\end{figure*}

When considering modal controllability for graph models constructed with Gaussian and FA weighting schemes, we observed that WRG, RL, WS, MD2, MD4, and MD8 all had small mean modal controllability values and low variance (Fig.~\ref{fig_m128n}A,C). RG had higher mean controllability values with larger variance and outliers, while BA had the highest mean values with small variance. For graph models constructed with the power-law weighting scheme, we observed relatively uniform behavior, all with low controllability values, low variance, and skewed right-tailed distributions toward higher controllability values (Fig.~\ref{fig_m128n}D). The exception was the BA graph, which had high average controllability and low variance. The SC-weighted graphs varied in mean value (with RL and WS having the lowest mean and BA having the highest mean) but had similar variance with slightly right-tailed distributions (Fig.~\ref{fig_m128n}B). One-way ANOVAs indicated significant effects of graph type for all of the four edge weight distributions (see Table~\ref{tab:anovas1}).

\begin{figure*}
	\begin{center}
		\centerline{\includegraphics[width=0.9\textwidth]{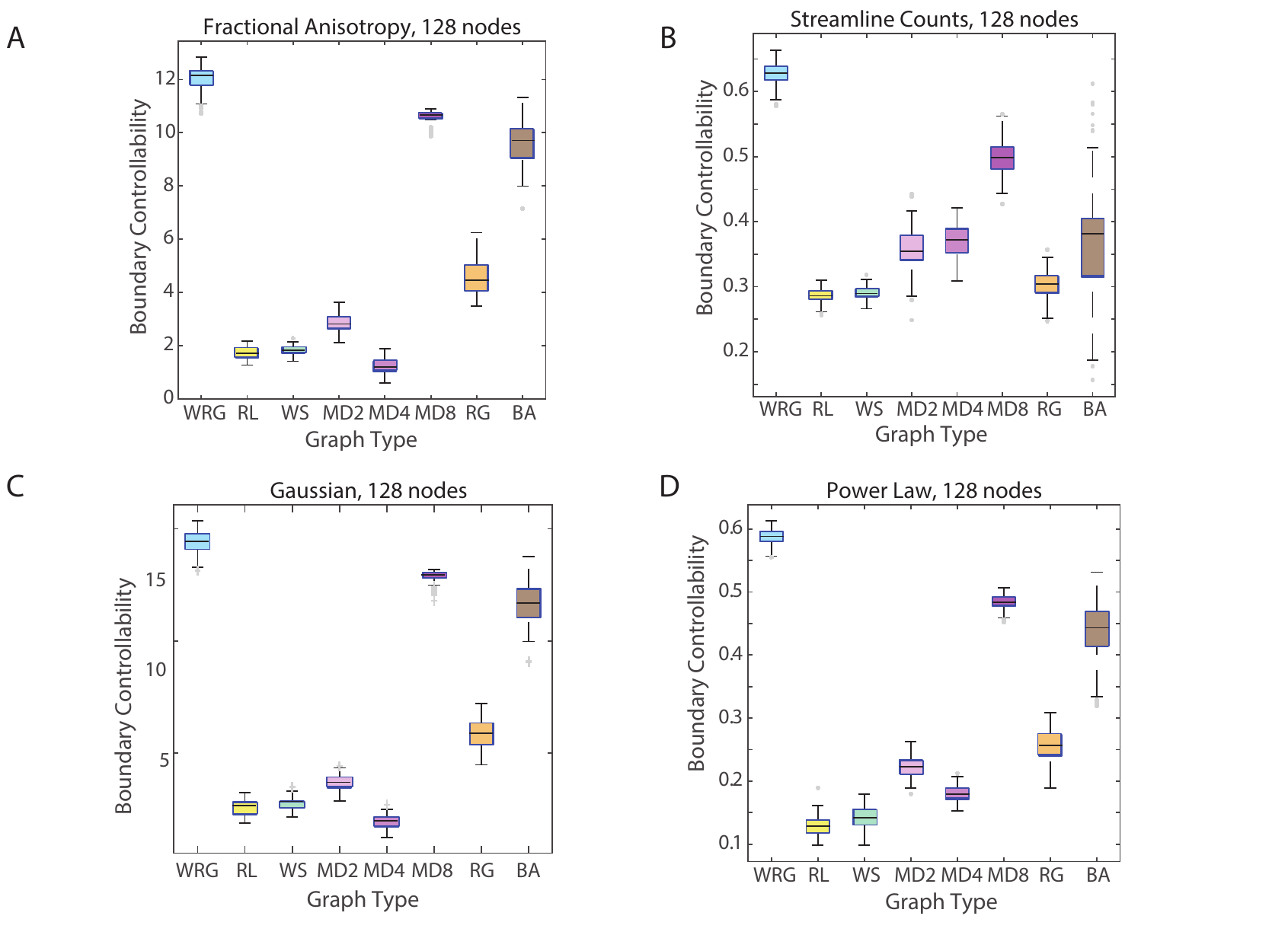}}
		\caption{\textbf{Variation in boundary controllability as a function of edge weighting and graph model.} Boundary controllability values were averaged over nodes in each graph model ensemble, and therefore boxplots show variation over graphs in the ensemble. Results are shown for four edge weighting schemes: \emph{(A)} FA, \emph{(B)} streamline counts, \emph{(C)} Gaussian, \emph{(D)} power-law. The eight graph models include the weighted random graph (WRG), the ring lattice (RL), the Watts-Strogatz small-world (WS), the modular graphs (MD2, MD4, MD8), the random geometric (RG), and the Barabasi-Albert preferential attachment (BA) models. Note that these results are presented for $\rho$ =  $10^{-8}$; we observe consistent results across thresholds of $10^{-5}$, $10^{-8}$, and $10^{-11}$ of edge weight distribution (see Supplement).} \label{fig_b128n}
		\vspace*{-10mm}
	\end{center}
\end{figure*}

When considering boundary controllability for graph models constructed with the Gaussian weighting scheme, we observed that RL, WS, MD2, and MD4 all had similarly small boundary controllability values and low variance (Fig.~\ref{fig_b128n}A). Then, RG and BA had increasingly higher controllability values and variance. MD8 had higher controllability values with small variance and a slightly left-tailed distribution of outliers. WRG had the highest controllability values and relatively small variance. For graph models constructed with the FA weighting scheme, trends were similar but the variance in the mean values for each type of graph tended to be higher and also more right-skewed compared to those of the Gaussian-weighted graphs (Fig.~\ref{fig_b128n}C). For graph models constructed with the SC-weighted graphs, we observed relatively uniform behavior, with low mean values (ranging from 0 to 1), low variance, and right-tailed distributions, with BA graph having the longest tail (Fig.~\ref{fig_b128n}B). The power law-weighted graphs had similar behavior to the SC-weighted graphs but had higher variance overall (Fig.~\ref{fig_b128n}D). One-way ANOVAs indicated significant effects of graph type for all of the four edge weight distributions (see Table~\ref{tab:anovas1}).

To summarize, when considering trends within a single edge weighting scheme, it is important to note that because edge weight distributions were exactly the same across each of the graph types, this guarantees that differences in controllability are due to network topology rather than the effects of differing edge weights. Since each of the graph types exhibits distinct behavior of controllability values for all types of controllability, this suggests that the topology of a network largely influences global, average, modal, and boundary controllability. Further, when considering trends across edge weighting schemes for a single graph model, it is important to note that this guarantees that differences in controllability are due to the effects of differing edge weights rather network topology. The similarity of trends in the controllability values between the FA and Gaussian weighting and then between the SC and power law weighting suggests that the nature of edge weights influences controllability when edge weight distributions are more normal (FA and Gaussian) \emph{versus} very skewed (SC and power law). \newline

\begin{table*}
 \centering
 \caption {\textbf{Effect of Graph Model on Controllability Statistics.} Results of one-way ANOVAs assessing the effect of graph model on each controllability statistic, and each edge weighting scheme, as shown in the boxplots in Fig.~\ref{fig_g128n}, ~\ref{fig_a128n}, ~\ref{fig_m128n}, and ~\ref{fig_b128n}. }
 \begin{tabular}{c c}
 \hline
~ & Global \\
 \hline
Fractional Anisotropy & $F(7) = 16.41$, $p=2.8 \times 10^{-20}$ \\
Gaussian & $F(7) = 23.92$, $p=1.3 \times 10^{-29}$ \\
Streamline Counts & $F(7)=1559.39$, $p=0.00$ \\
Power-law &	$F(7) = 671.62$, $p=0.00$ \\
\hline
~ & Average \\
\hline
Fractional Anisotropy & $F(7)=79446.27$, $p=0.00$ \\
Gaussian & $F(7)=71074.27$, $p=0.00$\\
Streamline Counts & $F(7)=760.92$, $p=0.00$ \\
Power-law & $F(7)=11748.73$, $p=0.00$ \\
\hline
~ & Modal \\
\hline
Fractional Anisotropy & $F(7)= 22830.9$, $p=0.00$ \\
Gaussian & $F(7)=21809.02$, $p=0.00$ \\
Streamline Counts & $F(7)=1525.89$, $p=0.00$ \\
Power-law & $F(7)=11765.4$, $p=0.00$ \\
\hline
~ & Boundary \\
\hline
Fractional Anisotropy & $F(7)=10015.69$, $p=0.00$ \\
Gaussian & $F(7)=10968.61$, $p=0.00$ \\
Streamline Counts & $F(7)=873.59$, $p=0.00$ \\
Power-law & $F(7)=5838.1$, $p=0.00$\\
 \hline
 \end{tabular}  \label{tab:anovas1}
\end{table*}

\subsection*{Relation between controllability statistics across graphs}

In prior work, we have observed that average and modal controllability, averaged over nodes, are positively related to one another across brain networks \cite{tang2016structural}. These data suggest that brain networks that are structurally predisposed to be effective in moving network dynamics into easy-to-reach states (via average control) are also the brain networks that are structurally predisposed to be effective in moving network dynamics into difficult-to-reach states (via modal control). Here, we ask whether this positive relationship between average and modal controllability across networks holds in canonical graph models.

\begin{figure*}
	\begin{center}
		\centerline{\includegraphics[width=0.9\textwidth]{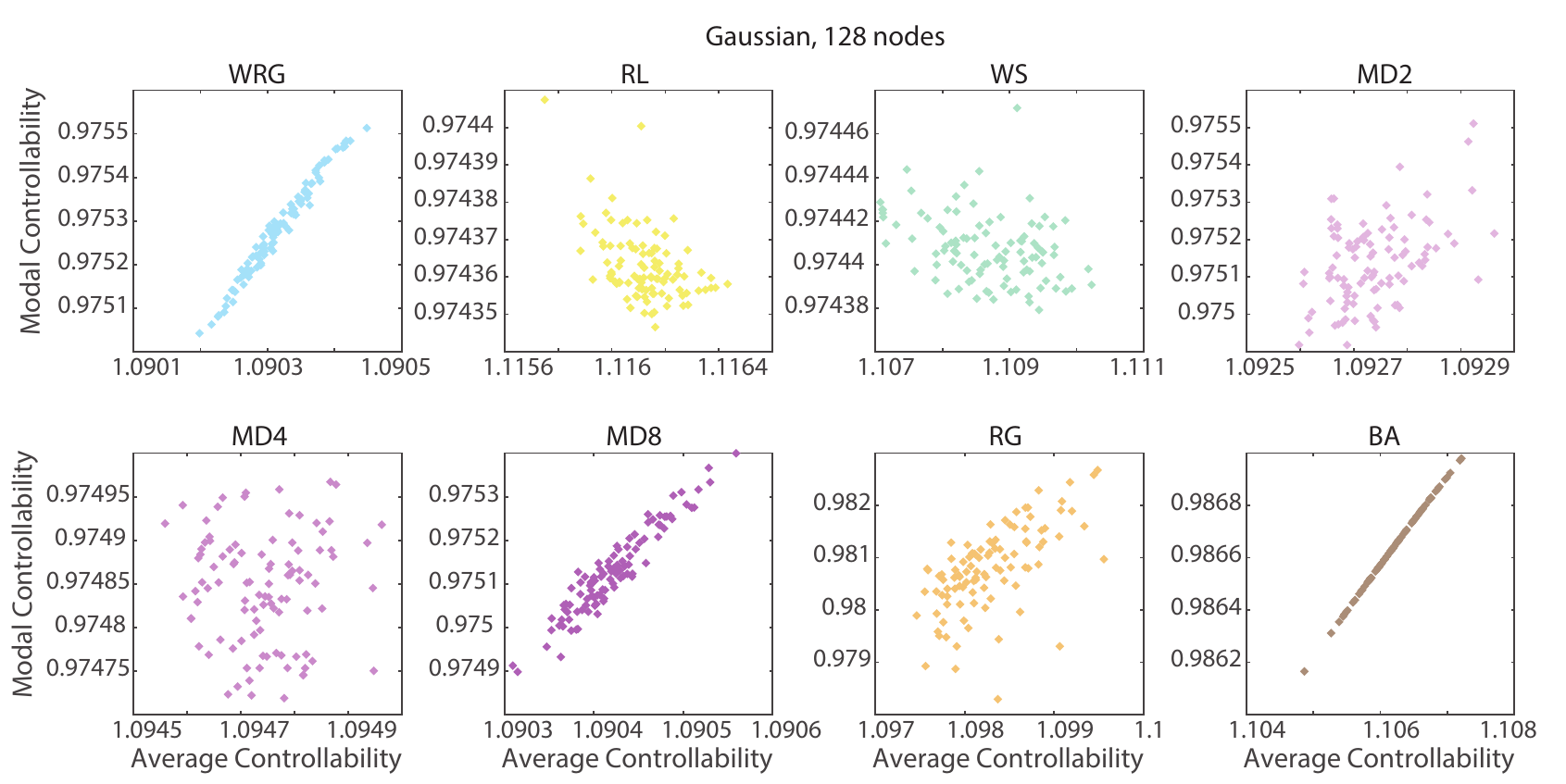}}
		\caption{\textbf{Relation between average and modal controllability across graphs for the Gaussian edge weighting scheme.} Average and modal controllability values were averaged over nodes in each graph model ensemble, and therefore scatterplots show values for each graph in the ensemble. Results are shown for the Gaussian edge weighting scheme. The eight graph models include the weighted random graph (WRG), the ring lattice (RL), the Watts-Strogatz small-world (WS), the modular graphs (MD2, MD4, MD8), the random geometric (RG), and the Barabasi-Albert preferential attachment (BA) models.  } \label{fig_mav128n}
		\vspace*{-10mm}
	\end{center}
\end{figure*}

In the Gaussian weighting scheme, we observed that one of the graphs (MD4) exhibited no correlation between average and modal controllability, while MD8 and RG had a moderate positive correlation, and WRG and BA had nearly linear positive correlations (Fig.~\ref{fig_mav128n}). Indeed, across other weighting schemes, there was no consistent positive or negative relation between average and modal controllability (see the Supplementary Materials). Specifically, in the FA weighting scheme, we observed that most graphs did not exhibit a strong correlation between average and modal controllability, except for WRG and BA, which had a moderate positive and strong positive correlation, respectively. Interestingly, in the streamline count weighting scheme, we observed that RL, WS, MD2, MD4, and MD8 exhibited a moderate negative correlation, while WRG and RG had weak negative correlations. BA exhibited a nearly linear, strong positive correlation. In the power-law weighting scheme, all 8 graph models exhibited a strong negative correlation between average and modal controllability that was nearly linear. For Spearman $rho$ correlation coefficients, see Table.~\ref{tab:spearman1}; we note that Gaussian and FA-weighted graphs were less likely to display a significant correlation between average and modal controllability than streamline-count and power-law weighted graphs.

\begin{table*}
 \centering
 \caption {\textbf{Relation between average and modal controllability across graphs.} Spearman $\rho$-values and corresponding $p$-values for the correlations between average and modal controllability statistics across graphs in an ensemble, after averaging nodal values across all nodes in each graph. Weighting schemes are abbreviated by ``FA'' (fractional anisotropy), ``Str'' (streamline counts), ``G'' (Gaussian), and ``PL'' (power-law). The $p$-values stated to be zero are simply estimated to be zero.}
\begin{tabular}{|c|c|c|c|c|c|c|c|c|}
\hline
\textbf{128 node} & WRG & RL & WS & MD2 & MD4 & MD8 & RG & BA \\
\hline
\hline
FA & ~ & ~ & ~ & ~ & ~ & ~ & ~ & ~ \\
\hline
$\rho$ & 0.8802 & -0.3996 & -0.3628 & -0.0723 & -0.1699 & 0.5382 & -0.1377 & 0.9999 \\
$p$ & 0 & 0 & 0.0002 & 0.4738 & 0.091 & 0 & 0.1717 & 0 \\
\hline
Str & ~ & ~ & ~ & ~ & ~ & ~ & ~ & ~ \\
\hline
$\rho$ & -0.3605 & -0.8782 & -0.898 & -0.7497 & -0.8715 & -0.5263 & -0.4028 & 0.9999 \\
$p$ & 0.0003 & 0 & 0 & 0 & 0 & 0 & 0 & 0 \\
\hline
G & ~ & ~ & ~ & ~ & ~ & ~ & ~ & ~ \\
\hline
$\rho$ & 0.9854 & -0.393 & -0.3614 & 0.3398 & -0.0003 & 0.9278 & 0.6545 & 0.9999 \\
$p$ & 0 & 0.0001 & 0.0002 & 0.0006 & 0.9981 & 0 & 0 & 0 \\
\hline
PL & ~ & ~ & ~ & ~ & ~ & ~ & ~ & ~ \\
\hline
$\rho$ & -1 & -1 & -1 & -1 & -1 & -1 & -1 & -1 \\
$p$ & 0 & 0 & 0 & 0 & 0 & 0 & 0 & 0 \\
\hline
\end{tabular} \label{tab:spearman1}
\end{table*}

\begin{figure*}
	\begin{center}
		\centerline{\includegraphics[width=0.9\textwidth]{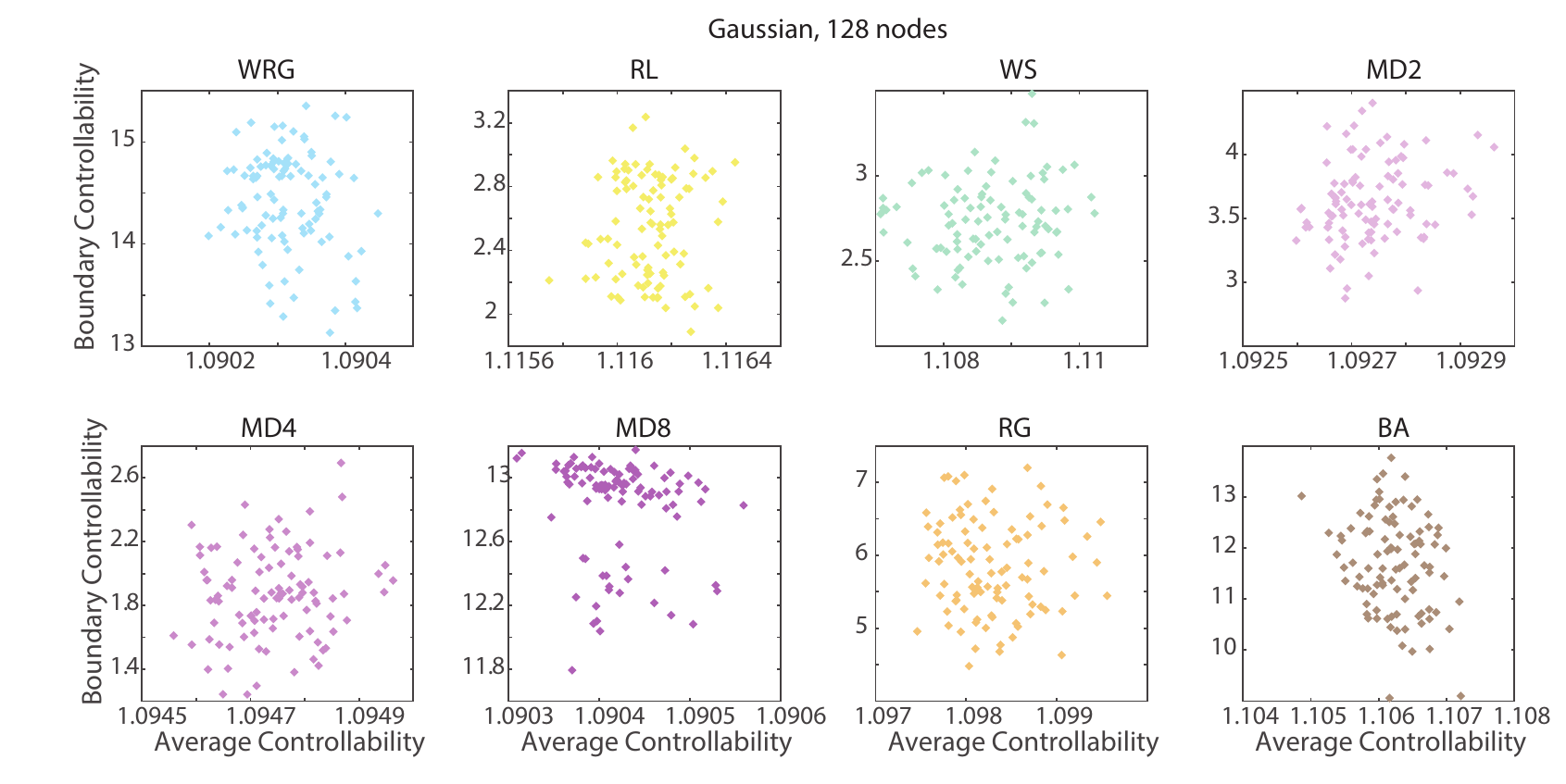}}
		\caption{\textbf{Relation between average and boundary controllability across graphs for the Gaussian edge weighting scheme.} Average and boundary controllability values were averaged over nodes in each graph model ensemble, and therefore scatterplots show values for each graph in the ensemble. Results are shown for the Gaussian edge weighting scheme. The eight graph models include the weighted random graph (WRG), the ring lattice (RL), the Watts-Strogatz small-world (WS), the modular graphs (MD2, MD4, MD8), the random geometric (RG), and the Barabasi-Albert preferential attachment (BA) models.  } \label{fig_bav128n}
		\vspace*{-10mm}
	\end{center}
\end{figure*}

\begin{figure*}
	\begin{center}
		\centerline{\includegraphics[width=0.9\textwidth]{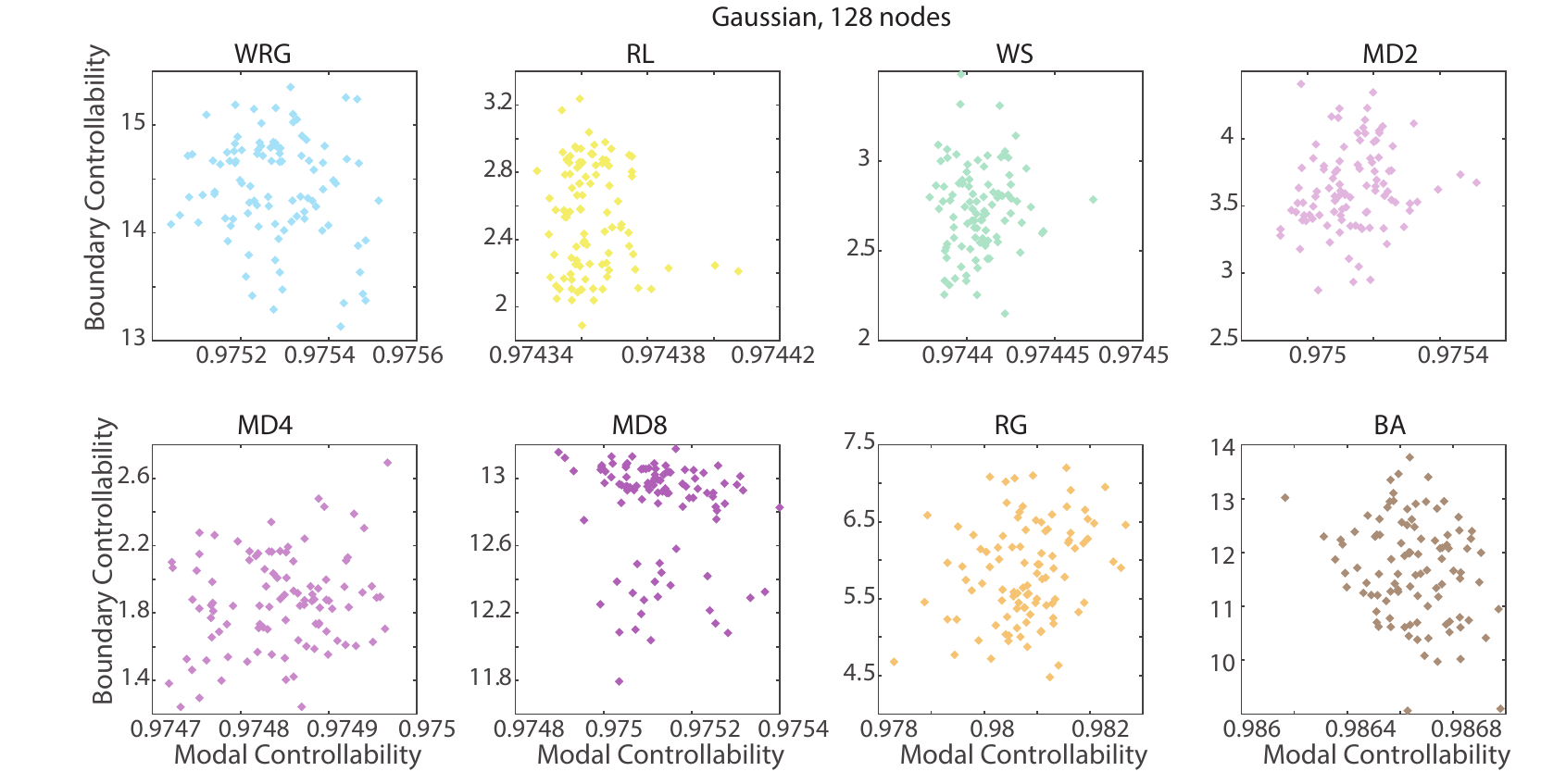}}
		\caption{\textbf{Relation between modal and boundary controllability across graphs for the Gaussian edge weighting scheme.} Modal and boundary controllability values were averaged over nodes in each graph model ensemble, and therefore scatterplots show values for each graph in the ensemble. Results are shown for the Gaussian edge weighting scheme. The eight graph models include the weighted random graph (WRG), the ring lattice (RL), the Watts-Strogatz small-world (WS), the modular graphs (MD2, MD4, MD8), the random geometric (RG), and the Barabasi-Albert preferential attachment (BA) models.  } \label{fig_bmv128n}
		\vspace*{-10mm}
	\end{center}
\end{figure*}

While not previously reported in prior work, we also asked whether boundary controllability statistics were correlated with either average or modal controllability statistics across graphs in an ensemble, after averaging nodal values across all nodes in each graph. In general for Gaussian weighting schemes, we observe that average and modal controllability do not tend to be strongly correlated with boundary controllability across graph instances (Fig.~\ref{fig_bav128n} and Fig.~\ref{fig_bmv128n}). Across other weighting schemes, we observe the same trends, for no strong relationships between boundary controllability statistics and either average or modal controllability statistics across graphs in an ensemble. (Table~\ref{tab:spearman1ba} and Table~\ref{tab:spearman1bm}).

\begin{table*}
 \centering
 \caption {\textbf{Relation between average and boundary controllability across graphs.} Spearman $\rho$-values and corresponding $p$-values for the correlations between average and boundary controllability statistics across graphs in an ensemble, after averaging nodal values across all nodes in each graph. Weighting schemes are abbreviated by ``FA'' (fractional anisotropy), ``Str'' (streamline counts), ``G'' (Gaussian), and ``PL'' (power-law). The $p$-values stated to be zero are simply estimated to be zero.}
\begin{tabular}{|c|c|c|c|c|c|c|c|c|}
\hline
\textbf{128 node} & WRG & RL & WS & MD2 & MD4 & MD8 & RG & BA \\
\hline
\hline
FA & ~ & ~ & ~ & ~ & ~ & ~ & ~ & ~ \\
\hline
$\rho$ & 0.0082 & -0.0487 & 0.0409 & 0.0591 & 0.1444 & -0.1797 & -0.3611 & 0.0226  \\
$p$ &  0.9350 & 0.6301 & 0.6856 & 0.5587 & 0.1514 & 0.0737 & 0.0002 & 0.8229 \\
\hline
Str & ~ & ~ & ~ & ~ & ~ & ~ & ~ & ~ \\
\hline
$\rho$ &  -0.1674 & 0.0844 & 0.0087 & -0.0218 & 0.0593 & -0.0462 & -0.0141 & 0.2487  \\
$p$ & 0.0959 & 0.4030 & 0.9316 & 0.8292 & 0.5571 & 0.6476 & 0.8887 & 0.0128 \\
\hline
G & ~ & ~ & ~ & ~ & ~ & ~ & ~ & ~ \\
\hline
$\rho$ &  -0.1036 & 0.0181 & 0.0713 & 0.2220 & 0.0797 & -0.3191 & -0.0922 &  -0.1824 \\
$p$ &  0.3047 & 0.8578 & 0.4804 & 0.0266 & 0.4300 & 0.0013 & 0.3611 & 0.0693 \\
\hline
PL & ~ & ~ & ~ & ~ & ~ & ~ & ~ & ~ \\
\hline
$\rho$ & 0.0015 & 0.0060 & 0.0253 & 0.0616 & 0.0527 & 0.0005 & 0.1025 & -0.2127 \\
$p$ &  0.9879 & 0.9525 & 0.8027 & 0.5422 & 0.6018 & 0.9964 & 0.3096 & 0.0338 \\
\hline
\end{tabular} \label{tab:spearman1ba}
\end{table*}

\begin{table*}
 \centering
 \caption {\textbf{Relation between modal and boundary controllability across graphs.} Spearman $\rho$-values and corresponding $p$-values for the correlations between modal and boundary controllability statistics across graphs in an ensemble, after averaging nodal values across all nodes in each graph. Weighting schemes are abbreviated by ``FA'' (fractional anisotropy), ``Str'' (streamline counts), ``G'' (Gaussian), and ``PL'' (power-law). The $p$-values stated to be zero are simply estimated to be zero.}
\begin{tabular}{|c|c|c|c|c|c|c|c|c|}
\hline
\textbf{128 node} & WRG & RL & WS & MD2 & MD4 & MD8 & RG & BA \\
\hline
\hline
FA & ~ & ~ & ~ & ~ & ~ & ~ & ~ & ~ \\
\hline
$\rho$ & 0.1187 & 0.1454 & -0.0179 & 0.2755 & 0.1499 & -0.2597 & 0.3055 & 0.0217  \\
$p$ &  0.2391 & 0.1488 & 0.8597 & 0.0057 & 0.1363 & 0.0092 & 0.0021 & 0.8303  \\
\hline
Str & ~ & ~ & ~ & ~ & ~ & ~ & ~ & ~ \\
\hline
$\rho$ &  0.0467 & -0.0093 & -0.0613 & 0.2103 & -0.0274 & 0.0542 & -0.1405 & 0.2500  \\
$p$ &  0.6443 & 0.9265 & 0.5437 & 0.0359 & 0.7864 & 0.5919 & 0.1631 & 0.0123  \\
\hline
G & ~ & ~ & ~ & ~ & ~ & ~ & ~ & ~ \\
\hline
$\rho$ &   -0.1040 & -0.0263 & 0.0758 & 0.2185 & 0.1310 & -0.2689 & 0.2230 &  -0.1806 \\
$p$ &  0.3026 & 0.7948 & 0.4528 & 0.0292 & 0.1935 & 0.0070 & 0.0259 & 0.0723 \\
\hline
PL & ~ & ~ & ~ & ~ & ~ & ~ & ~ & ~ \\
\hline
$\rho$ &  -0.0014 & -0.0060 & -0.0256 & -0.0616 & -0.0527 & -0.0005 & -0.1025 & 0.2121 \\
$p$ &   0.9889 & 0.9525 & 0.8001 & 0.5422 & 0.6018 & 0.9961 & 0.3096 & 0.0343 \\
\hline
\end{tabular} \label{tab:spearman1bm}
\end{table*}

\subsection*{Effect of graph size on nodal variation in network controllability statistics across graph models}

To assess the reliability and reproducibility of our results, we next examined the impact of graph size ($n = 128$, $256$, or $512$) on network controllability statistics, and their modulation by graph model for the Gaussian edge weight distribution (see the Supplementary Materials). We observed that global, average, modal, and boundary controllability values for graphs of 256 and 512 nodes largely maintained the same trends as for graphs with 128 nodes. However, mean global, average, and modal controllability values tended to decrease with increasing graph size (consistent with analytical studies), while mean boundary controllability values tended to increase with increasing graph size (with the exception of MD8). Variance of average and modal controllability tended to decrease with increasing graph size. Importantly, because size varied within the same type of controllability for the same set of graphs, this guarantees that differences in controllability are due to the effects of differing network size rather than network topology. The similarity of trends across graph sizes suggests that the controllability properties are maintained in networks of different size but may be accentuated through decreased spread or mean value with increasing size. \newline

\subsection*{Nodal variation in network controllability statistics across graph models}

In the previous section, we examined how network controllability statistics varied over graphs within a given ensemble, and whether that variation differed as a function of graph model. These questions focused on values of controllability statistics that were calculated at each node of the network separately, and then averaged over nodes in the network. In this section, we turn to an examination of the nodal variation in network controllability statistics, and ask questions regarding how nodal variation differs across graph models, and between controllability statistics. Following the procedure outlined in the Methods section, for each controllability type we took the mean of the 128 sorted nodal controllability values across all 100 graphs in a given model ensemble, giving us 128 controllability values averaged over the graph instances. For each controllability type, we used four identical edge weight distributions corresponding to fractional anisotropy (FA) and streamline counts (SC) from real brain data, Gaussian distribution, and power law distribution.

\begin{figure*}
	\begin{center}
		\centerline{\includegraphics[width=0.9\textwidth]{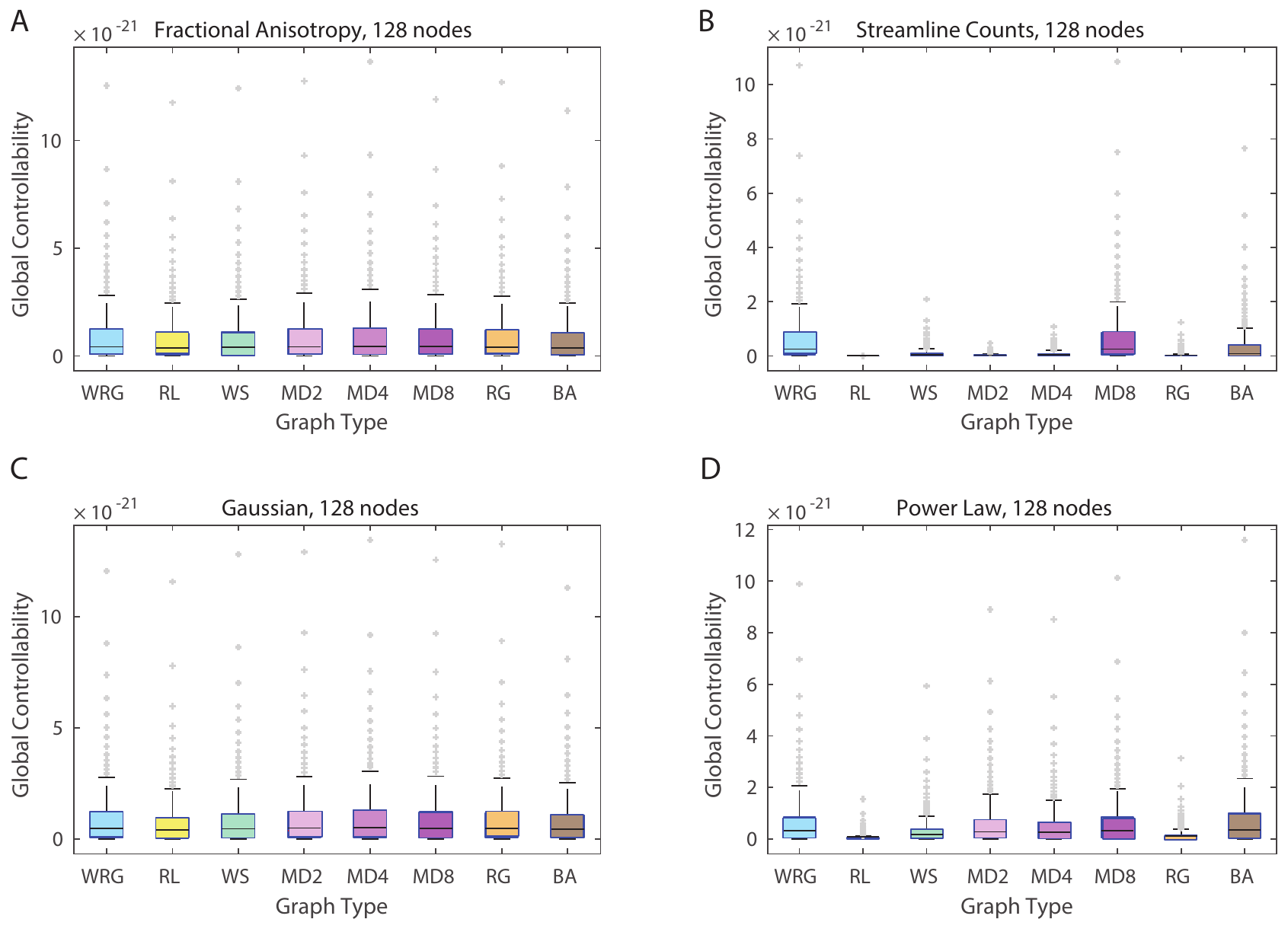}}
		\caption{\textbf{Nodal variation in global controllability as a function of edge weighting and graph model.} Global controllability values were averaged over instances in each graph model ensemble, and therefore boxplots show variation over nodes in the graph. Results are shown for four edge weighting schemes: \emph{(A)} FA, \emph{(B)} streamline counts, \emph{(C)} Gaussian, \emph{(D)} power-law. The eight graph models include the weighted random graph (WRG), the ring lattice (RL), the Watts-Strogatz small-world (WS), the modular graphs (MD2, MD4, MD8), the random geometric (RG), and the Barabasi-Albert preferential attachment (BA) models. } \label{fig_g128}
		\vspace*{-10mm}
	\end{center}
\end{figure*}

When considering global controllability for graph models constructed with Gaussian and FA weighting schemes, we observed that all graph models had nearly identical right-tailed distributions with primarily low global controllability values and low variance (Fig.~\ref{fig_g128}A,C). For graph models constructed with steamline counts and power-law weighting schemes, we observed a similar pattern but with generally lower variance for graphs such as RL, WS, and RG (as well as MD2 and MD4 for the streamline count weighting) and slightly less skew toward higher values (Fig.~\ref{fig_g128}B,D). One-way ANOVAs indicated significant effects of graph type for all of the four edge weight distributions (see Table~\ref{tab:anovas2}).

\begin{figure*}
	\begin{center}
		\centerline{\includegraphics[width=0.9\textwidth]{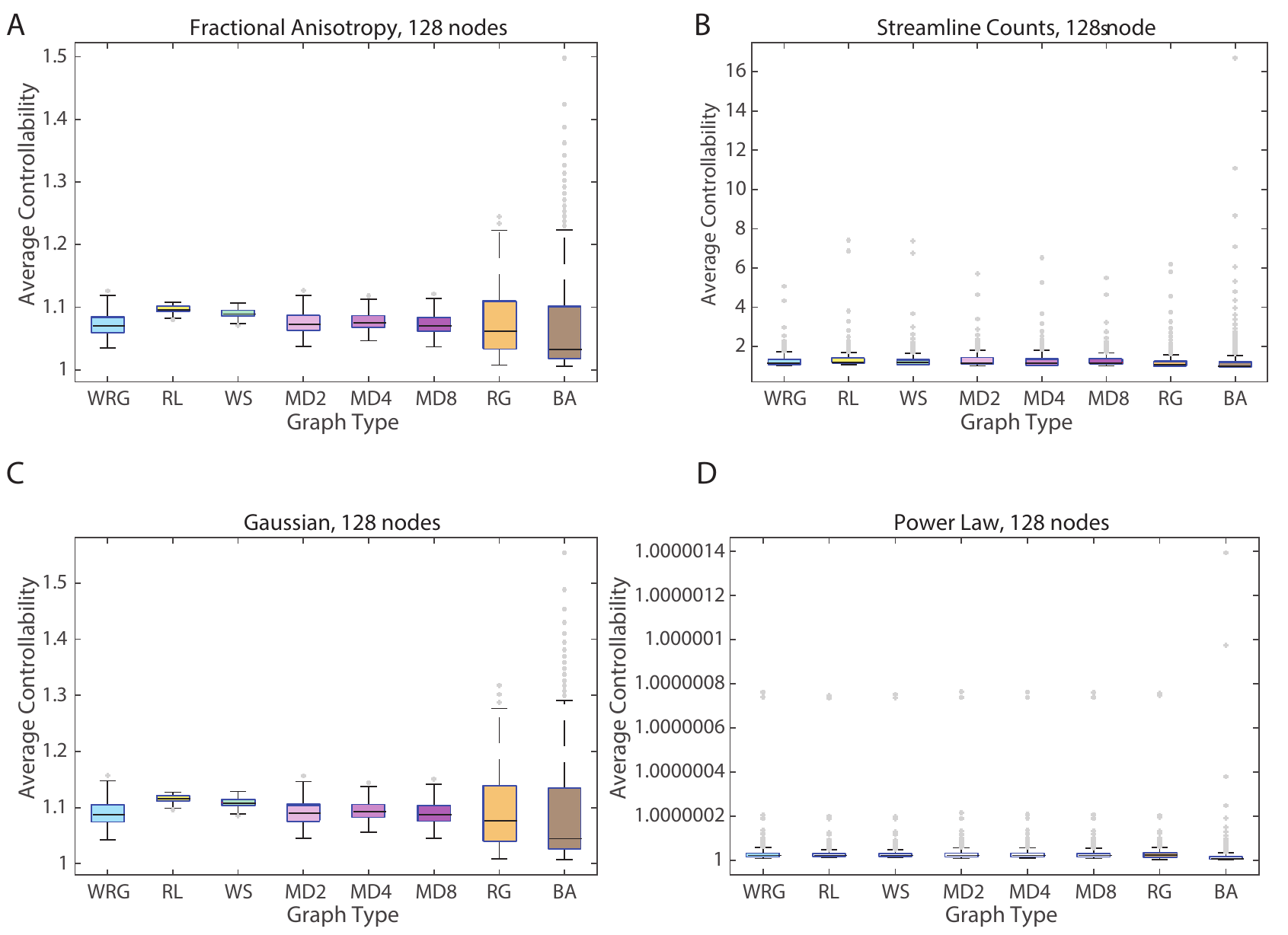}}
		\caption{\textbf{Nodal variation in average controllability as a function of edge weighting and graph model.} Average controllability values were averaged over instances in each graph model ensemble, and therefore boxplots show variation over nodes in the graph. Results are shown for four edge weighting schemes: \emph{(A)} FA, \emph{(B)} streamline counts, \emph{(C)} Gaussian, \emph{(D)} power-law. The eight graph models include the weighted random graph (WRG), the ring lattice (RL), the Watts-Strogatz small-world (WS), the modular graphs (MD2, MD4, MD8), the random geometric (RG), and the Barabasi-Albert preferential attachment (BA) models. } \label{fig_a128}
		\vspace*{-10mm}
	\end{center}
\end{figure*}

When considering average controllability for graph models constructed with Gaussian and FA weighting schemes, we observed that WRG, MD2, MD4, and MD8 all had similar mean average controllability values with low variance (Fig.~\ref{fig_a128}A,C). RL and WS had slightly higher mean average controllability values and the lowest variance of all the graph types. RG and BA had the lowest mean average controllability values and highest variance, and the BA model was also most skewed, with a significant right-tailed distribution toward higher controllability values. In contrast, the graphs constructed with power-law and streamline count weighting schemes exhibited relatively uniform behavior, all with low controllability values, low variance, and skewed right-tailed distributions toward higher controllability values (Fig.~\ref{fig_a128}B,D). One-way ANOVAs indicated significant effects of graph type for three of the four edge weight distributions: FA, streamline counts, and Gaussian but not power-law (see Table~\ref{tab:anovas2}).

\begin{figure*}
	\begin{center}
		\centerline{\includegraphics[width=0.9\textwidth]{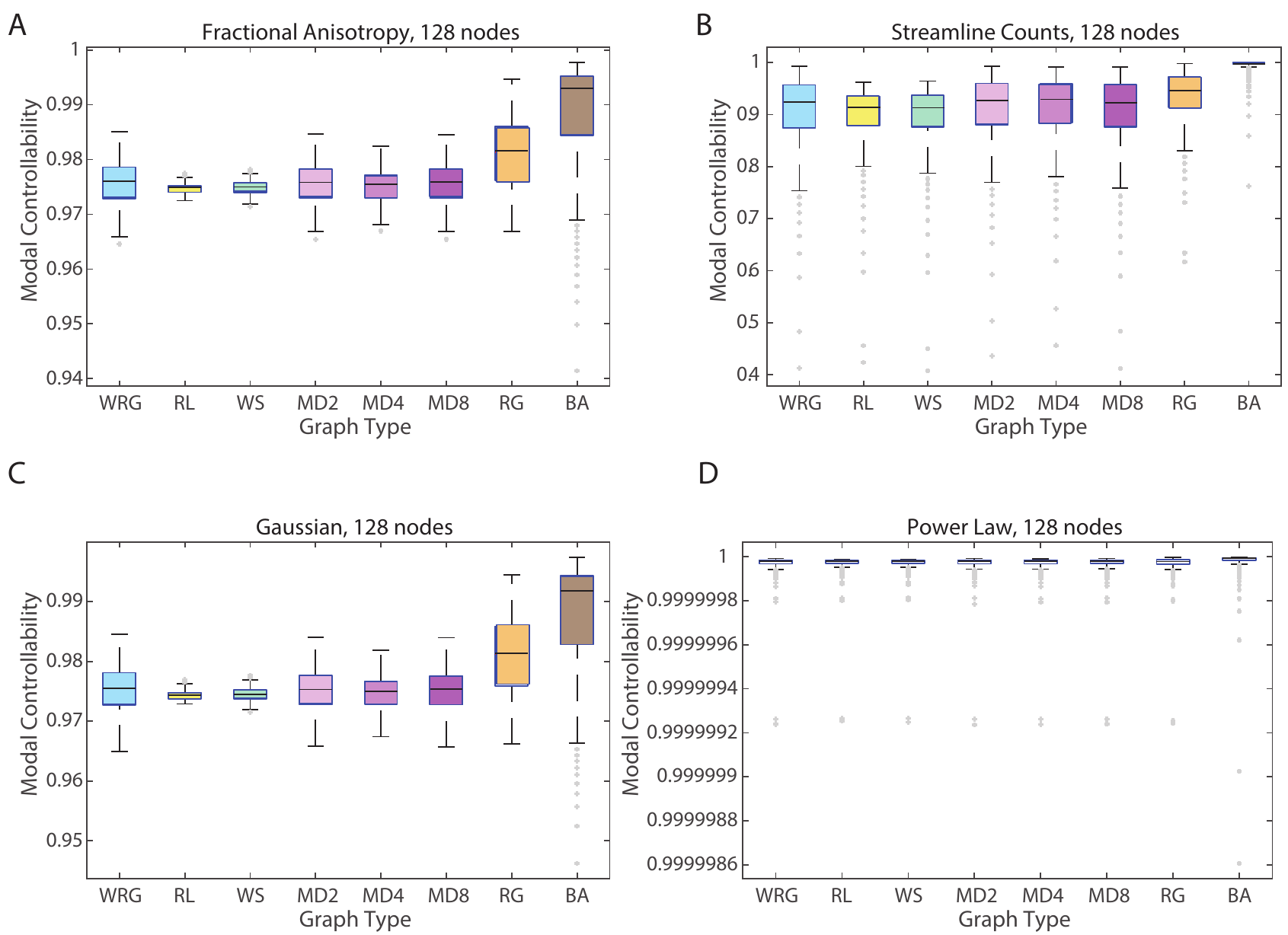}}
		\caption{\textbf{Nodal variation in modal controllability as a function of edge weighting and graph model.} Modal controllability values were averaged over instances in each graph model ensemble, and therefore boxplots show variation over nodes in the graph. Results are shown for four edge weighting schemes: \emph{(A)} FA, \emph{(B)} streamline counts, \emph{(C)} Gaussian, \emph{(D)} power-law. The eight graph models include the weighted random graph (WRG), the ring lattice (RL), the Watts-Strogatz small-world (WS), the modular graphs (MD2, MD4, MD8), the random geometric (RG), and the Barabasi-Albert preferential attachment (BA) models.  } \label{fig_m128}
		\vspace*{-10mm}
	\end{center}
\end{figure*}

When considering the modal controllability for graph models constructed with Gaussian and FA weighting schemes, we observed that WRG, RL, WS, MD2, MD4, and MD8 all had similar mean modal controllability values and similar variance (Fig.~\ref{fig_m128}A,C). RG had a higher mean controllability value, while BA had the highest. RL and WS had the smallest variance, while RG and BA had the largest variance. The BA model was also the most skewed, with a significant left-tailed distribution toward lower controllability values. In contrast, the graphs constructed with power-law and streamline count weighting schemes exhibited nearly uniform behavior, all with high controllability values, low variance (BA model lowest), and skewed left-tailed distributions toward lower controllability values (Fig.~\ref{fig_m128}B,D). The power law-weighted graphs had lower variance and slightly higher mean than those of the graphs weighted by streamline counts. One-way ANOVAs indicated significant effects of graph type for three of the four edge weight distributions: FA, streamline counts, and Gaussian but not power-law (see Table~\ref{tab:anovas2}).

Finally, when considering boundary controllability for graph models constructed with a Gaussian weighting schemes, we observed that RL, WS, MD2, MD4, RG, and BA had right-tailed distributions, with increasing variance and skewness from RL to WS, MD2, RG, and BA. MD4 was also right-skewed and had the lowest variance. WRG and MD8 were not skewed toward higher controllability values, had more symmetric distributions, and had lower variance than the other graph types. The boundary controllability values of the graph types with FA weighting and power law weighting followed similar trends to those observed with the Gaussian weighting. The differences were that for power law weighting, the overall controllability values were an order of magnitude smaller than those of Gaussian weighting and the MD4 graph had higher variance. For FA weighting, RL, WS, and MD4 all had lower variance than for Gaussian weighting. The graphs weighted by streamline counts exhibited a different trend, with low means (all between 0 and 1), low variance, and slight right-tailed distributions, with the BA graph being most skewed. One-way ANOVAs indicated significant effects of graph type for all of the four edge weight distributions (see Table~\ref{tab:anovas2}).

\begin{figure*}
	\begin{center}
		\centerline{\includegraphics[width=0.9\textwidth]{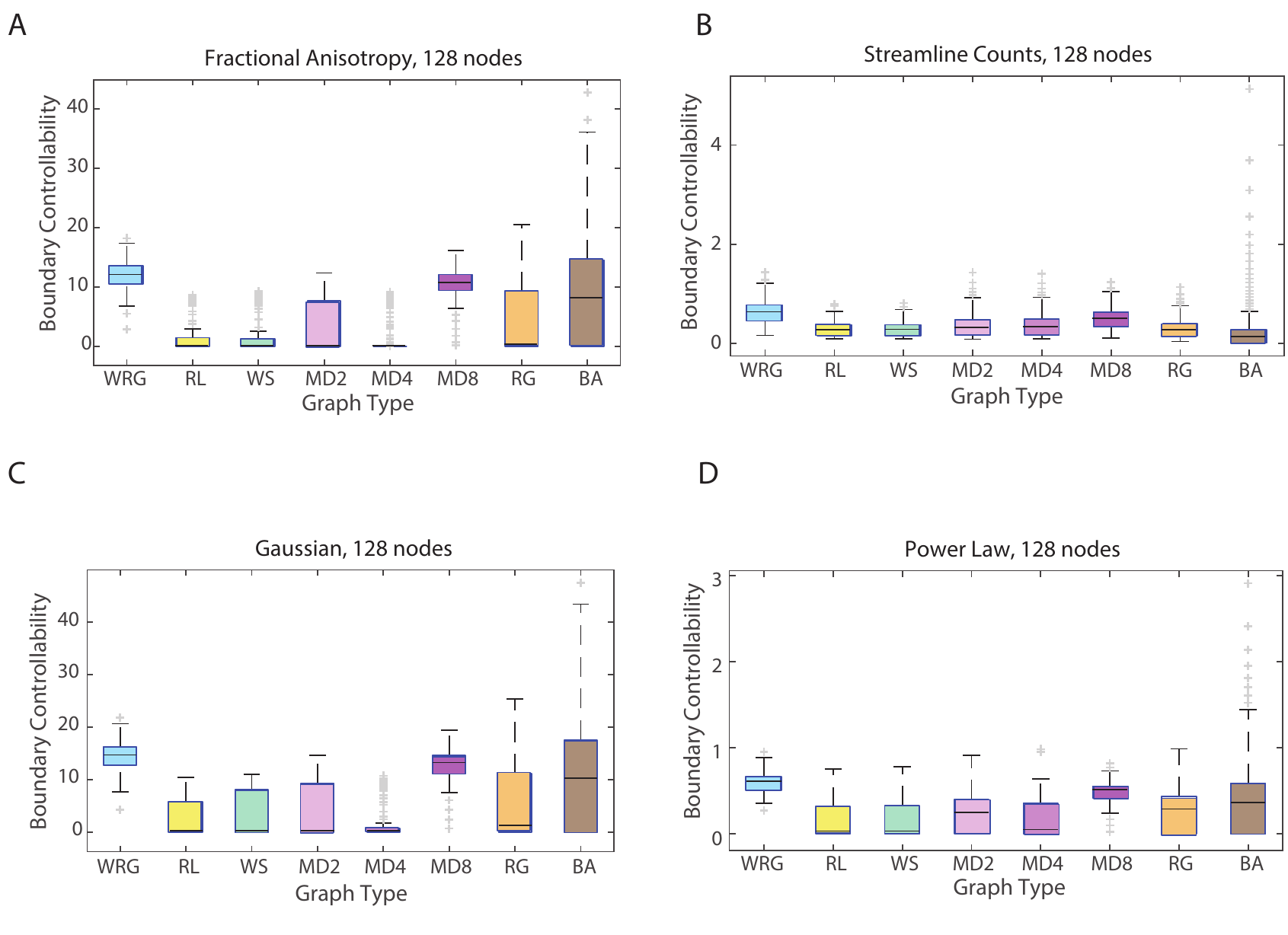}}
		\caption{\textbf{Nodal variation in boundary controllability as a function of edge weighting and graph model.} Boundary controllability values were averaged over instances in each graph model ensemble, and therefore boxplots show variation over nodes in the graph. Results are shown for four edge weighting schemes: \emph{(A)} FA, \emph{(B)} streamline counts, \emph{(C)} Gaussian, \emph{(D)} power-law. The eight graph models include the weighted random graph (WRG), the ring lattice (RL), the Watts-Strogatz small-world (WS), the modular graphs (MD2, MD4, MD8), the random geometric (RG), and the Barabasi-Albert preferential attachment (BA) models. Note that these results are presented for $\rho$ =  $10^{-8}$; we observe consistent results across percentiles of $10^{-5}$, $10^{-8}$, and $10^{-11}$ of edge weight distribution (see Supplement). } \label{fig_b128}
		\vspace*{-10mm}
	\end{center}
\end{figure*}

To summarize, when considering trends within a single edge weighting scheme, it is important to note that because edge weight distributions were exactly the same across each of the graph types, this guarantees that differences in controllability are due to network topology rather than the effects of differing edge weights. Since each of the graph types exhibits distinct behavior of controllability values for all types of controllability except average and modal controllability in power-law weighted graphs, this suggests that the topology of a network largely influences global, average, modal, and boundary controllability, but can be obfuscated in graphs whose edge weights follow a power-law distribution. \newline

\begin{table*}
 \centering
 \caption {\textbf{Effect of Graph Model on Nodal Controllability Statistics.} Results of one-way ANOVAs assessing the effect of graph model on each controllability statistic, and each edge weighting scheme, as shown in the boxplots in Fig.~\ref{fig_g128}, ~\ref{fig_a128}, ~\ref{fig_m128}, and ~\ref{fig_b128}. }
 \begin{tabular}{c c}
 \hline
~ & Global \\
 \hline
 Fractional Anisotropy & $F(7) =16.08$, $p=3.6 \times 10^{-20}$ \\
 Gaussian & $F(7)=23.31$, $p=1.8 \times 10^{-29}$ \\
 Streamline Counts & $F(7)=851.46$, $p=0.00$ \\
 Power-law &	$F(7) = 651.08$, $p=0.00$ \\
 \hline
~ & Average \\
 \hline
 Fractional Anisotropy & $F(7)=8.42$, $p=4.9 \times 10^{-10}$ \\
 Gaussian & $F(7) = 6.99$, $p=3.7 \times 10^{-8}$  \\
 Streamline Counts & $F(7)=7.36$, $p=1.2 \times 10^{-8}$ \\
 Power-law & $F(7)=0$, $p=1.00$ \\
  \hline
~ & Modal \\
 \hline
 Fractional Anisotropy & $F(7)=189.11$, $p=3.9 \times 10^{-179}$ \\
 Gaussian & $F(7)=182.97$, $p=4.6 \times 10^{-175}$ \\
 Streamline Counts & $F(7) = 1426.98$, $p=0.00$ \\
 Power-law & $F(7)=0$, $p=1.00$ \\
  \hline
~ & Boundary \\
 \hline
 Fractional Anisotropy & $F(7)=237.7$, $p=6 \times 10^{-209}$ \\
 Gaussian & $F(7)=253.1$, $p=1 \times 10^{-217}$\\
 Streamline Counts & $F(7)= 48.9$, $p = 5 \times 10^{-60}$ \\
 Power-law & $F(7)=149.57$, $p=2 \times 10^{-151}$\\
 \hline
 \end{tabular}  \label{tab:anovas2}
\end{table*}

\subsection*{Relation between controllability statistics across nodes}

In prior work, we have observed that average and modal controllability are inversely related to one another across regions in brain networks, both in non-invasive human neuroimaging data acquired in youth and adults \cite{gu2015controllability,tang2016structural} and in tract-tracing data acquire from macaque monkeys \cite{gu2015controllability}. These data suggest that regions of the brain that are structurally predisposed to be effective in moving the brain into easy-to-reach states (via average control) are different from the regions of the brain that are structurally predisposed to be effective in moving the brain into difficult-to-reach states (via modal control). Here, we ask whether this inverse relationship between average and modal controllability across nodes holds in canonical graph models.

\begin{figure*}
	\begin{center}
		\centerline{\includegraphics[width=0.9\textwidth]{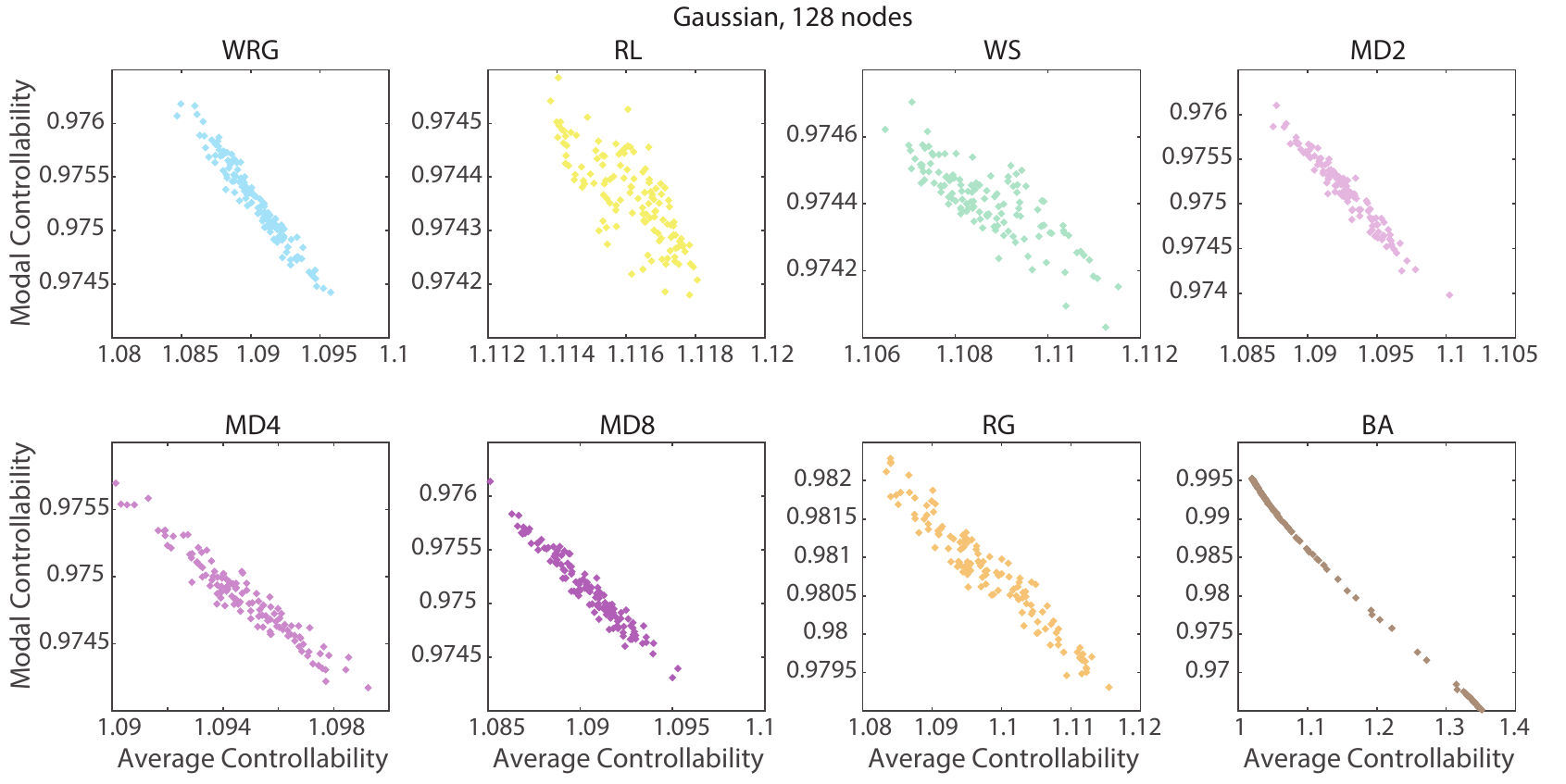}}
		\caption{\textbf{Relation between average and modal controllability across nodes for the Gaussian edge weighting scheme.} Average and modal controllability values were averaged over instances in each graph model ensemble, and therefore scatterplots show values for each node in the graph. Results are shown for the Gaussian edge weighting scheme. The eight graph models include the weighted random graph (WRG), the ring lattice (RL), the Watts-Strogatz small-world (WS), the modular graphs (MD2, MD4, MD8), the random geometric (RG), and the Barabasi-Albert preferential attachment (BA) models.  } \label{fig_mav128}
		\vspace*{-10mm}
	\end{center}
\end{figure*}

For graphs constructed with the Gaussian weighting scheme, we observed that average and modal controllability were negatively correlated with one another across all 8 graph models, with RL having the weakest and BA having the strongest correlation (Fig.~\ref{fig_mav128}). We observed similar trends in the graphs constructed with the FA weighting scheme, while in graphs constructed with the streamline count weighting scheme we observed the BA graph model to have the strongest correlation, and in the power-law weighting scheme we observed that the relationship between average and modal controllability was nearly perfectly linear across all graph models (see the Supplementary Materials). For Spearman $rho$ correlation coefficients, see Table.~\ref{tab:spearman2}.

\begin{table*}
 \centering
 \caption {\textbf{Relation between average and modal controllability across nodes.} Spearman $\rho$-values and corresponding $p$-values for the correlations between average and modal controllability statistics across nodes in a graph, after averaging values across graph. Weighting schemes are abbreviated by ``FA'' (fractional anisotropy), ``Str'' (streamline counts), ``G'' (Gaussian), and ``PL'' (power-law). The $p$-values stated to be zero are simply estimated to be zero.}
\begin{tabular}{|c|c|c|c|c|c|c|c|c|}
\hline
\textbf{128 node} & WRG & RL & WS & MD2 & MD4 & MD8 & RG & BA \\
\hline
\hline
FA & ~ & ~ & ~ & ~ & ~ & ~ & ~ & ~ \\
\hline
$\rho$ & -0.9716 & -0.7519 & -0.7967 & -0.9658 & -0.9462 & -0.9614 & -0.9578 & -0.9997 \\
$p$ & 0 & 0 & 0 & 0 & 0 & 0 & 0 & 0 \\
\hline
Str & ~ & ~ & ~ & ~ & ~ & ~ & ~ & ~ \\
\hline
$\rho$ & -0.8902 & -0.8306 & -0.8637 & -0.8152 & -0.7789 & -0.8304 & -0.7487 & -0.9995 \\
$p$ & 0 & 0 & 0 & 0 & 0 & 0 & 0 & 0 \\
\hline
G & ~ & ~ & ~ & ~ & ~ & ~ & ~ & ~ \\
\hline
$\rho$ & -0.9756 & -0.7764 & -0.7951 & -0.9702 & -0.9464 & -0.9683 & -0.9579 & -0.9996 \\
$p$ & 0 & 0 & 0 & 0 & 0 & 0 & 0 & 0 \\
\hline
PL & ~ & ~ & ~ & ~ & ~ & ~ & ~ & ~ \\
\hline
$\rho$ & -1 & -1 & -1 & -1 & -1 & -1 & -1 & -1 \\
$p$ & 0 & 0 & 0 & 0 & 0 & 0 & 0 & 0 \\
\hline
\end{tabular} \label{tab:spearman2}
\end{table*}

\begin{figure*}
	\begin{center}
		\centerline{\includegraphics[width=0.9\textwidth]{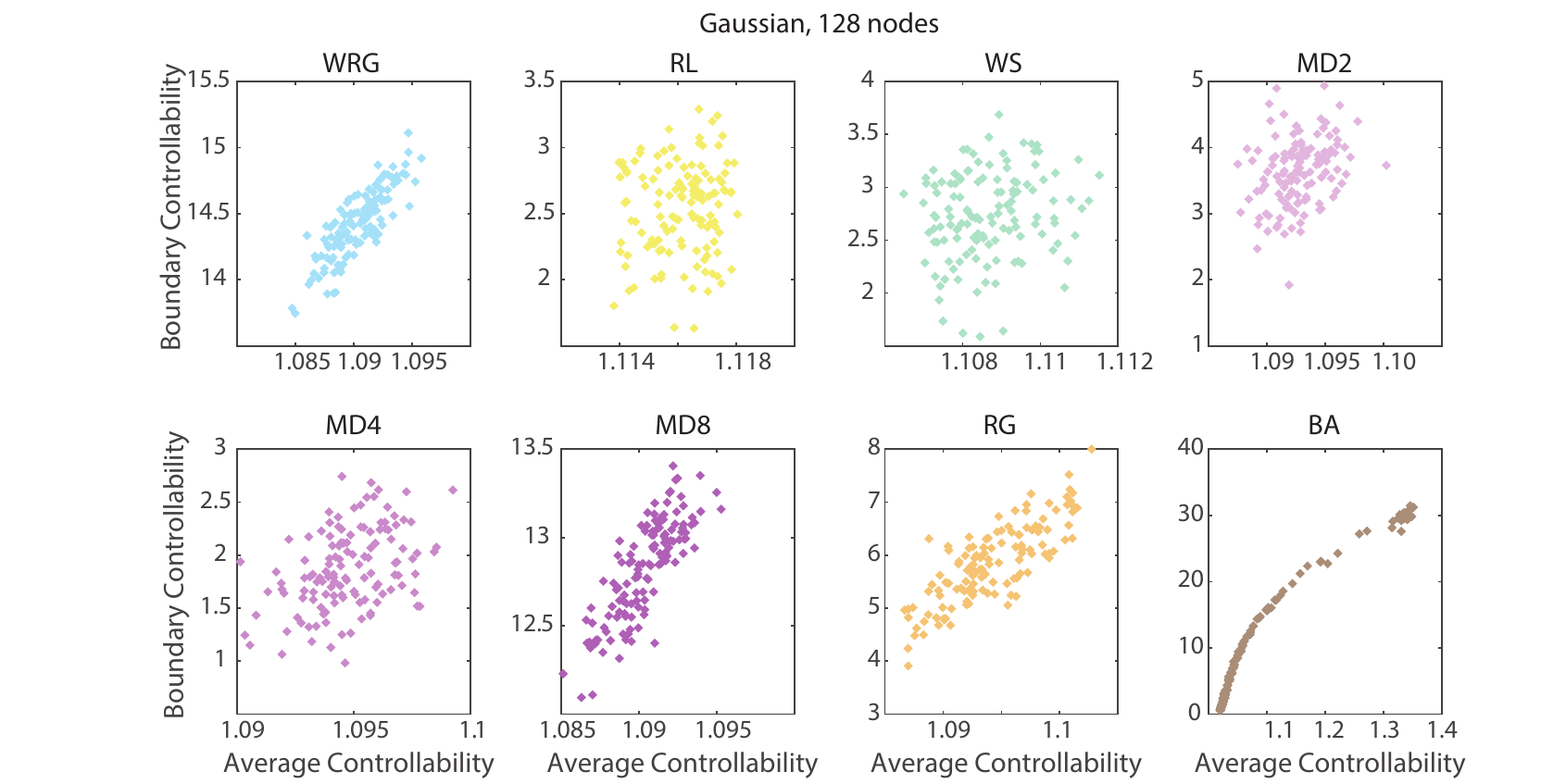}}
		\caption{\textbf{Relation between average and boundary controllability across nodes for the Gaussian edge weighting scheme.} Average and boundary controllability values were averaged over instances in each graph model ensemble, and therefore scatterplots show values for each node in the graph. Results are shown for the Gaussian edge weighting scheme. The eight graph models include the weighted random graph (WRG), the ring lattice (RL), the Watts-Strogatz small-world (WS), the modular graphs (MD2, MD4, MD8), the random geometric (RG), and the Barabasi-Albert preferential attachment (BA) models.  } \label{fig_bav128}
		\vspace*{-10mm}
	\end{center}
\end{figure*}

\begin{figure*}
	\begin{center}
		\centerline{\includegraphics[width=0.9\textwidth]{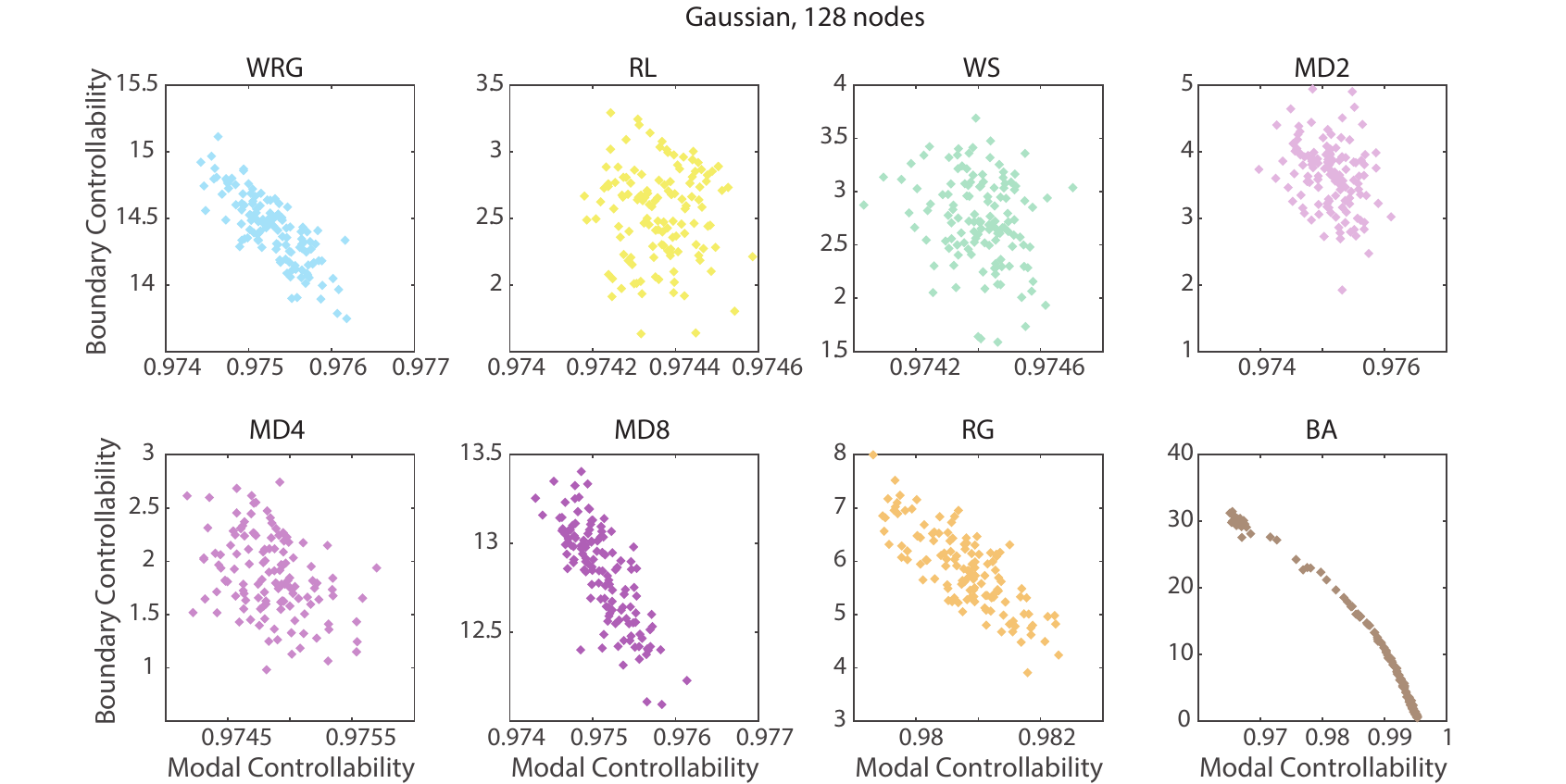}}
		\caption{\textbf{Relation between modal and boundary controllability across nodes for the Gaussian edge weighting scheme.} Modal and boundary controllability values were averaged over instances in each graph model ensemble, and therefore scatterplots show values for each node in the graph. Results are shown for the Gaussian edge weighting scheme. The eight graph models include the weighted random graph (WRG), the ring lattice (RL), the Watts-Strogatz small-world (WS), the modular graphs (MD2, MD4, MD8), the random geometric (RG), and the Barabasi-Albert preferential attachment (BA) models.  } \label{fig_bmv128}
		\vspace*{-10mm}
	\end{center}
\end{figure*}

In prior work, we have observed that average and boundary controllability are not strongly related to one another across regions in adult human brain networks \cite{gu2015controllability}. These data suggest that regions of the brain that are structurally predisposed to be effective in moving the brain into easy-to-reach states (via average control) may sometimes (but not consistently) also be the regions of the brain that are structurally predisposed to be effective in gating across network communities (via boundary control). Here, we ask whether the lack of a relationship between average and boundary controllability across nodes holds in canonical graph models. In general for Gaussian weighting schemes, we observe that average controllability tends to be positively correlated with boundary controllability across nodes (Fig.~\ref{fig_bav128}); in contrast, boundary controllability tends to be negatively correlated with modal controllability across nodes (Fig.~\ref{fig_bmv128}). These relationships appear to be least salient in small-world and modular graphs, which may explain why they were not previously observed in brain networks \cite{gu2015controllability}. Across other weighting schemes, we observe consistent trends, for average controllability to be positively related to boundary controllability, and for modal controllability to be negatively related to boundary controllability across nodes in a graph (Table~\ref{tab:spearman2ba} and Table~\ref{tab:spearman2bm}).

\begin{table*}
 \centering
 \caption {\textbf{Relation between average and boundary controllability across nodes.} Spearman $\rho$-values and corresponding $p$-values for the correlations between average and boundary controllability statistics across nodes in a graph, after averaging values across graph. Weighting schemes are abbreviated by ``FA'' (fractional anisotropy), ``Str'' (streamline counts), ``G'' (Gaussian), and ``PL'' (power-law). The $p$-values stated to be zero are simply estimated to be zero.}
\begin{tabular}{|c|c|c|c|c|c|c|c|c|}
\hline
\textbf{128 node} & WRG & RL & WS & MD2 & MD4 & MD8 & RG & BA \\
\hline
\hline
FA & ~ & ~ & ~ & ~ & ~ & ~ & ~ & ~ \\
\hline
$\rho$ &  0.7915 & 0.1495 & 0.1246 & 0.3393 & 0.4141 & 0.7464 & 0.7816 & 0.9939 \\
$p$ &  0 & 0.0921 & 0.1609 & 0.0001 & 0.0000 & 0 & 0 & 0 \\
\hline
Str & ~ & ~ & ~ & ~ & ~ & ~ & ~ & ~ \\
\hline
$\rho$ & 0.3324 & -0.0475 & -0.1106 & 0.3404 & 0.2763 & -0.0041 & 0.1169 & 0.9935 \\
$p$ & 0.0001 & 0.5944 & 0.2136 & 0.0001 & 0.0017 & 0.9634 & 0.1887 & 0 \\
\hline
G & ~ & ~ & ~ & ~ & ~ & ~ & ~ & ~ \\
\hline
$\rho$ &  0.8217 & 0.0665 & 0.1727 & 0.3213 & 0.4041 & 0.8181 & 0.7871 &  0.9960 \\
$p$ &  0 & 0.4551 & 0.0513 & 0.0002 & 0.0000 & 0 & 0 &  0 \\
\hline
PL & ~ & ~ & ~ & ~ & ~ & ~ & ~ & ~ \\
\hline
$\rho$ & 0.3400 & 0.5281 & 0.5678 & 0.4121 & 0.5852 & -0.1224 & 0.3145 & 0.9970 \\
$p$ &  0.0001 & 0.0000 & 0 & 0.0000 & 0 & 0.1683 & 0.0003 & 0 \\
\hline
\end{tabular} \label{tab:spearman2ba}
\end{table*}

\begin{table*}
 \centering
 \caption {\textbf{Relation between modal and boundary controllability across nodes.} Spearman $\rho$-values and corresponding $p$-values for the correlations between modal and boundary controllability statistics across nodes in a graph, after averaging values across graph. Weighting schemes are abbreviated by ``FA'' (fractional anisotropy), ``Str'' (streamline counts), ``G'' (Gaussian), and ``PL'' (power-law). The $p$-values stated to be zero are simply estimated to be zero.}
\begin{tabular}{|c|c|c|c|c|c|c|c|c|}
\hline
\textbf{128 node} & WRG & RL & WS & MD2 & MD4 & MD8 & RG & BA \\
\hline
\hline
FA & ~ & ~ & ~ & ~ & ~ & ~ & ~ & ~ \\
\hline
$\rho$ & -0.7499 & -0.1514 & -0.1795 & -0.3215 & -0.3864 & -0.6756 & -0.7584 & -0.9938  \\
$p$ &  0 &  0.0881 & 0.0428 & 0.0002 & 0.0000 & 0 & 0 & 0 \\
\hline
Str & ~ & ~ & ~ & ~ & ~ & ~ & ~ & ~ \\
\hline
$\rho$ & -0.4204 & -0.0180 & 0.1091 & -0.3065 & -0.3100 & -0.1203 & -0.0902 & -0.9940  \\
$p$ & 0.0000 & 0.8401 & 0.2201 & 0.0005 & 0.0004 & 0.1761 & 0.3111 & 0  \\
\hline
G & ~ & ~ & ~ & ~ & ~ & ~ & ~ & ~ \\
\hline
$\rho$ &  -0.8019 & -0.0056 & -0.2321 & -0.3100 & -0.3735 & -0.7761 & -0.7604 &  -0.9963  \\
$p$ & 0 & 0.9495 & 0.0085 & 0.0004 & 0.0000 & 0 & 0 &  0  \\
\hline
PL & ~ & ~ & ~ & ~ & ~ & ~ & ~ & ~ \\
\hline
$\rho$ &  -0.3400 & -0.5281 & -0.5678 & -0.4121 & -0.5852 & 0.1224 & -0.3145 & -0.9970\\
$p$ &  0.0001 & 0.0000 & 0 & 0.0000 & 0 & 0.1683 & 0.0003 & 0\\
\hline
\end{tabular} \label{tab:spearman2bm}
\end{table*}

\subsection*{Effect of graph size on nodal variation in network controllability statistics across graph models}

To assess the reliability and reproducibility of our results, we next examined the impact of graph size (n = 128, 256, or 512) on network controllability statistics, and their modulation by graph model for the Gaussian edge weight distribution (see the Supplementary Materials). We observed that global, average, modal, and boundary controllability values for graphs of 256 and 512 nodes largely maintained the same trends as for graphs with 128 nodes. However, mean global controllability values tended to decrease with increasing graph size, and variance of average, modal, and boundary controllability values tended to decrease with increasing graph size. In addition, for the MD8 graph, the mean boundary controllability decreases to close to 0 in the graph of size 512. Importantly, because size varied within the same type of controllability for the same set of graphs, this guarantees that differences in controllability are due to the effects of differing network size rather than network topology. The similarity of trends across graph sizes suggests that the controllability properties are maintained in networks of different size but may be accentuated through decreased spread or mean value with increasing size. \newline	

\subsection*{Similarities in patterns of controllability statistics between graph models}

\begin{figure*}
	\begin{center}
		\centerline{\includegraphics[width=0.9\textwidth]{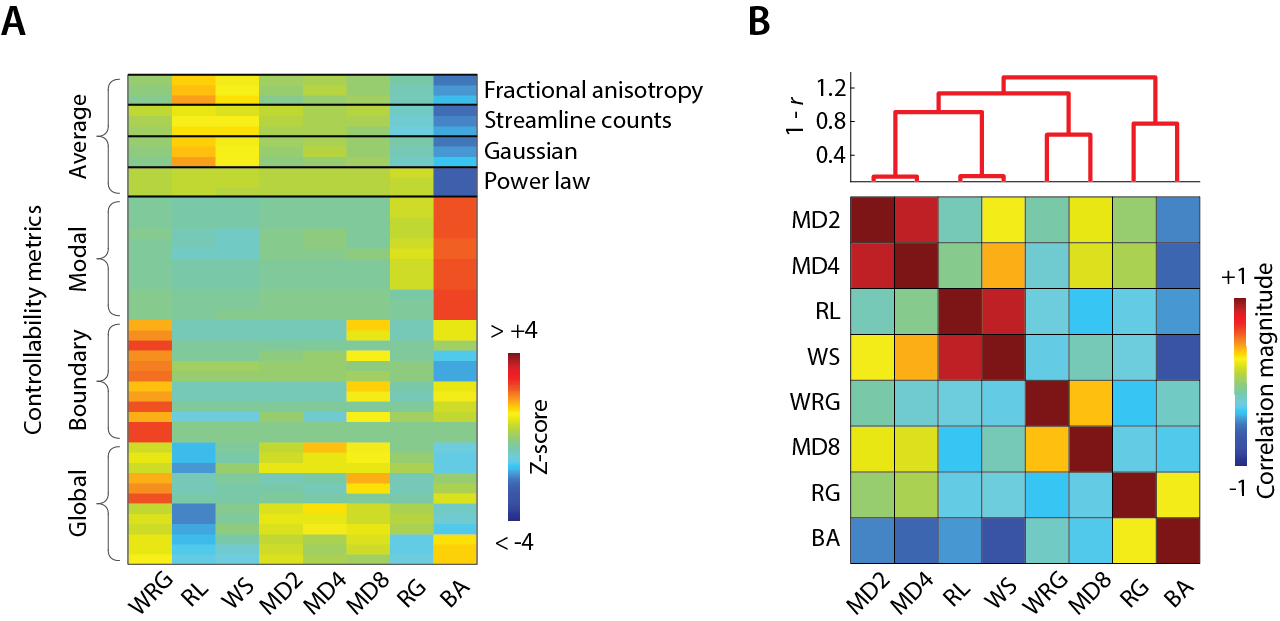}}
		\caption{\textbf{Clustering controllability features across graph models.} For each graph model, we extracted the median average, modal, boundary, and global controllability statistics for the four edge weight distributions and graphs of 128, 256, and 512 nodes. \emph{(A)} Matrix of controllability statistics $z$-scored across graph models so that each row has a mean of zero and unit variance. \emph{(B)} We computed the Pearson correlation coefficient, $r$, of controllability metrics for every pair of graphs and clustered that matrix, revealing graph types whose controllability statistics were correlated across the manipulations studied in this report.} \label{correlatedFeatures}
		\vspace*{-10mm}
	\end{center}
\end{figure*}

Finally, we asked whether certain graphs with similar topologies might show similar patterns of controllability statistics across edge weighting schemes and network sizes. To address this question we treated average, modal, boundary, and global controllability as features of interest, and extracted their median values for each of the eight graph models. Rather than express these statistics as raw scores which can sometimes differ by many orders of magnitude, we standardized them across graph models and expressed them as $z$-scores. We repeated this process for all combinations of edge weight distributions and graph sizes, resulting in 48 features for each graph model (Fig.~\ref{correlatedFeatures}A).

To assess the similarity of controllability statistics across graph models, we computed the model-by-model correlation matrix of $z$-scored features (Fig.~\ref{correlatedFeatures}B). This matrix was marked by heterogeneity. Intuitively, we observed that networks with similar topological features exhibited similar behavior in terms of their controllability statistics. For example, the two- and four-module graphs, which both featured large, segregated modules, displayed patterns of statistics that were highly correlated with one another ($r = 0.86$, $p < 10^{-14}$). Likewise, the ring lattice and the Watts-Strogatz graphs, which featured highly regular organization, displayed patterns of statistics that were also highly correlated ($r = 0.85$, $p < 10{-14}$). Not surprisingly, the Barab\'{a}si-Albert model, which was the only model we included that had a heavy-tailed degree (and strength) distribution, was dissimilar to the other models, on average. These findings suggest that network topological properties induce similarities in the behavior of controllability statistics across graph models.

\section*{Discussion}

The representation of complex systems as graphs or networks has proven useful in a wide variety of domains for the study of physical \cite{papadopoulos2016evolution}, technological \cite{tan2014traffic}, biological \cite{steinway2015inference,baldassano2016topological}, ecological \cite{proulx2005network}, and social \cite{ilany2016social,evans2016multiple} processes. While initial work in each of these domains focused on developing descriptive statistics to characterize the nature of the system's graph representation, more recent efforts have turned to developing predictions of system function and fundamental theories of system dynamics \cite{bassett2017network}. One particularly powerful approach to both prediction and theory for these systems comes from the emerging field of \emph{network control}, which provides analytical results for the response of a network system to internal or external perturbation \cite{bassett2017emerging}. The application of these tools to neural systems has recently provided important insights into possible mechanisms of cognitive control \cite{gu2015controllability}, energetic explanations for baseline activation of the default mode system \cite{betzel2016optimally}, and the emergence of diverse dynamics over the course of normative neurodevelopment \cite{tang2016structural}.

Despite these initial successes, a basic understanding of the performance of these tools on graphs with different topologies or geometries is lacking \cite{bianchin2015role}. Here we address this challenge by studying commonly-applied network controllability statistics assuming linear system dynamics to several canonical graph models built with distinct edge weight distributions. We find that both graph topology and edge weight distribution can impact network controllability statistics estimated at single regions or across the whole network. These data underscore the importance of assessing network controllability statistics in one's own data (as well as statistic-statistic relations) rather than relying on assumptions built from other data. More importantly, the results point to the necessity of developing analytical descriptions of the relations between topology, geometry, and control.

\subsection*{Understanding topological drivers of control}

Intuitively, one might imagine that the topology of a graph should have a non-trivial impact on the types of control strategies that the system can perform or respond to. A star graph, containing many nodes only (and directly) connected to a single central node, may have quite different responses to energy injected into the central node than to energy injected into the peripheral nodes \cite{pasqualetti2014isotropic}. Moreover, both of these responses may be quite distinct from the response of a lattice graph to local perturbations at any of the nodes \cite{bianchin2015role}. The intuition that topology matters for control is one that has now been supported by decades of prior literature \cite{reinschke1988multivariable}, and is quantitatively demonstrated in our results, which show that canonical graph models tend to display significantly different values for global, average, modal, and boundary controllability, both in variations across a graph ensemble and in variations across network nodes. This intuition has further motivated both the development of controllability statistics \cite{pasqualetti2014controllability}, and of intuitions for how controllability may relate to the system's dynamics (synchronizability) and geometry (symmetry) \cite{whalen2015observability}. Recent work has pinpointed specific features of edge weight vectors associated with each node that explain the energy expected for control \cite{kim2016topological}, and this work also suggested that these features vary in a meaningful way over species whose origins span evolutionary time scales.

Collectively, these studies motivate the question of whether neural systems have a characteristic graph topology that supports their specific functions as information processing systems in organisms. Canonical graph models with simple rules specifying the existence or growth of connections have proven useful in initial forays into this question \cite{klimm2014resolving,samu2014influence}. While evidence suggests that region location \cite{kaiser2006nonoptimal}, spatial embedding \cite{samu2014influence}, and mechanisms for growth \cite{klimm2014resolving} are important drivers of graph topology in neural systems, most graph models offer reasonable explanations for only one or a very few characteristic features of neural networks. Canonical graph models are therefore commonly used as benchmarks against which to compare real-world topologies, rather than as exact replicas of the biological system under study. As benchmarks, it is important to understand the expected controllability profiles of these graph models -- as we do here -- so as to inform interpretations regarding the biological specificity of control profiles observed empirically. Indeed, our observations will be useful in determining the degree to which simple connection rules (including random, small-world, and preferential attachment) can account for controllability statistics in neural systems.

\subsection*{Nontrivial impact of edge weight distributions on controllability of weighted networks}

An important contribution of our work stems from the fact that we do not study binary graph models, but instead examine graphs that have been weighted by drawing from both statistical and empirical functional forms for edge weight distributions. Our choice to focus on weighted graph models was motivated by recent work demonstrating that assessing binary versions of weighted graphs can provide inaccurate intuitions regarding a network's architecture, and by extension its function \cite{bassett2016small}. For example, quite dense graphs can appear to lack small-world architecture if studied as a binary matrix, but display strong small-world architecture when edge weights are taken into account \cite{muldoon2016small}. This discrepancy can be understood when considering the fact that weighting schemes tell us about the geometry (weight distribution, and weight location) of a network \cite{bassett2012influence}. Importantly, edge weights can have a direct impact on the potential to control, and on the energy required for control of neural systems \cite{kim2016topological}. Here, we observe that edge weight distributions can either enhance or obfuscate differences in controllability profiles across graph models, either in variations across a graph ensemble and in variations across network nodes. These findings suggest that it is wise to be cautious about inferring the controllability profile of a graph model independent of knowledge regarding its edge weight distribution. They also suggest interesting future directions for network design, particularly in cases where the graph architecture is fixed by external constraints but where the edge weights can be varied with the goal of enhancing or decrementing control.

\subsection*{Relations between controllability statistics prescribed by graph topology}

Initial efforts focused on a narrow class of graphs showed that average and modal controllability were positively correlated with one another across graphs instances \cite{tang2016structural}, and negatively correlated with one another across nodes \cite{gu2015controllability}, while average and boundary controllability were not significantly correlated with one another across nodes. Here we show that only one of these observations holds true across both graph models and edge weight distributions: that average and modal controllability are correlated with one another across nodes. In contrast to prior work in brain graphs, we show that average and modal controllability can be positively, negatively, or non-significantly correlated with one another across graph instances in an ensemble, and that boundary controllability tends to be positively correlated with average controllability (and negatively correlated with modal controllability) across nodes in a graph.  These data suggest that the relationship between controllability statistics depends strongly on the graph model's topological architecture, and on the observed edge weight distribution. These findings are interesting because they suggest the possibility of designing networks with different strengths for one type of control versus another, or for specific relationships between control profiles. Such a possibility is further bolstered by the fact that we observe in a hierarchical clustering procedure that certain graph models share greater similarity in their entire profile of controllability statistics (across the dimensions of size, edge weight distribution, etc. studied here) to some graph models than to other graph models.

\subsection*{Methodological considerations}

There are several methodological considerations that are pertinent to this work. First, we note that the network controllability statistics that we study are based fundamentally on a linear model of dynamics \cite{pasqualetti2014controllability}. Such a model is clearly appropriate for linear systems, but its application to systems characterized by nonlinear dynamics must be considered carefully. Practically speaking, linear models can provide excellent predictive power for a system in the neighborhood of the operating point \cite{leith2000survey}. In the context of neural systems, linear models of dynamics have proven useful both at the ensemble and large-scale regional levels in predicting intrinsic dynamics \cite{galan2008network,honey2009predicting}. Moreover, linear predictions of response to control input have been validated in nonlinear models of cortical columns \cite{muldoon2016stimulation}. These studies support the investigation of linear control in neural systems, but do not preclude future studies of explicitly nonlinear control \cite{cornelius2013realistic,motter2015network}, which could also prove useful in understanding neural systems \cite{tang2017control}.

A second important consideration relates to the control strategies that we study: global, average, modal, and boundary controllability. While these notions have proven useful both in man-made \cite{pasqualetti2014controllability} and natural \cite{gu2015controllability} systems, nevertheless, it is intuitively plausible that other as-yet-undefined control strategies may also prove relevant. Indeed, the definition of control metrics for complex networks is a fairly new area of research \cite{pasqualetti2014controllability,betzel2016optimally}, and future work is likely to develop a wider battery of statistics.

A third important consideration relates to limitations of the data that we used to construct our empirical edge weight distributions. Diffusion imaging is a powerful non-invasive neuroimaging technique\cite{weigell2000fiber}, which has only recently become commonplace in the construction of human (and non-human) connectomes \cite{hagmann2010mr,johansenberg2009using}. The technique is relatively new, and tractography algorithms applied to the data continue to be refined \cite{jbabdi2011tractography,pestilli2014evaluation}. It will be important in future work to evaluate other empirically-estimated edge weight distributions as they become available.

A fourth point is that we chose a specific normalization factor for our matrices to ensure stability. However, we note that different choices of normalization may accentuate \emph{versus} de-emphasize different scales of dynamics, and it will be interesting in future to study how the choice of normalization impacts observed patterns of controllability.

A fifth point is that we study controllability from a single node only. However, methods do exist for studying multi-point control \cite{betzel2016optimally}, and an important future direction for research is to understand how graph models differ in their capacities for multi-point control. Here, our focus on single node control is justified because we are interested in node variability, and not on specific controllability values per say.

A sixth point is that the computation of the controllability Gramian is intrinsically difficult because its smallest eigenvalue is in fact very small. Here we report the estimated numerical values and statistics, and only interpret differences in ensembles of graphs, and not differences between pairs of graphs.

A final important consideration relates to the set of graph models that we study. While we cover many of the canonical models that are frequently studied in network science, and especially those more frequently studied in the context of human and non-human brain networks \cite{klimm2014resolving,sizemore2017classification}, the set is in no way exhaustive. It would be interesting in future work to further develop biologically-motivated growth models that may more accurately take into account the neurophysiological processes of cell migration, synaptic plasticity and pruning, and other mutually trophic influences on neural development.

\section*{Conclusion}

In conclusion, we here examine a set of statistics that characterize diverse control strategies of complex networked systems whose dynamics can be approximated by a linear, noise-free, discrete-time, and time-invariant model. We apply these statistics to graph models whose edge weights are drawn from both empirical and statistically-defined functional forms. We show that controllability metrics, and their relations to one another, differ across graph models, and that those relations can be either elucidated or clouded by the distribution of edge weights in the graph. We observe that modular graph models show the most positively correlated patterns of controllability values across network size, controllability statistic, and edge weight distribution, while the Watts-Strogatz small-world model and Barabasi-Albert preferential attachment model show the most negatively correlated patterns. Our study offers intuitions for how controllability statistics behave in common graph models used as benchmarks for studies of brain networks in both human and non-human species. More generally, it suggests interesting future directions in designing networks to display a pattern of controllability statistics (and relations between them), particularly when the graph architecture is fixed by external constraints but the edge weights can be varied to enhance or decrement control.

\section*{Acknowledgments}
The authors would like to acknowledge support from the John D. and Catherine T. MacArthur Foundation, the Alfred P. Sloan Foundation, the Army Research Laboratory and the Army Research Office through contract numbers W911NF-10-2-0022 and W911NF-14-1-0679, the National Institute of Health (2-R01-DC-009209-11, 1R01HD086888-01, R01-MH107235, R01-MH107703, R01MH109520, 1R01NS099348 and R21-M MH-106799), the Office of Naval Research, and the National Science Foundation (BCS-1441502, CAREER PHY-1554488, BCS-1631550, and CNS-1626008).The content is solely the responsibility of the authors and does not necessarily represent the official views of any of the funding agencies.

\clearpage
\newpage
\clearpage
\newpage
\section*{Supporting Materials}

\subsection*{Supplementary Results}

\subsubsection*{Relation between average and modal controllability across graphs}

\begin{figure*}
	\begin{center}
		\centerline{\includegraphics[width=0.9\textwidth]{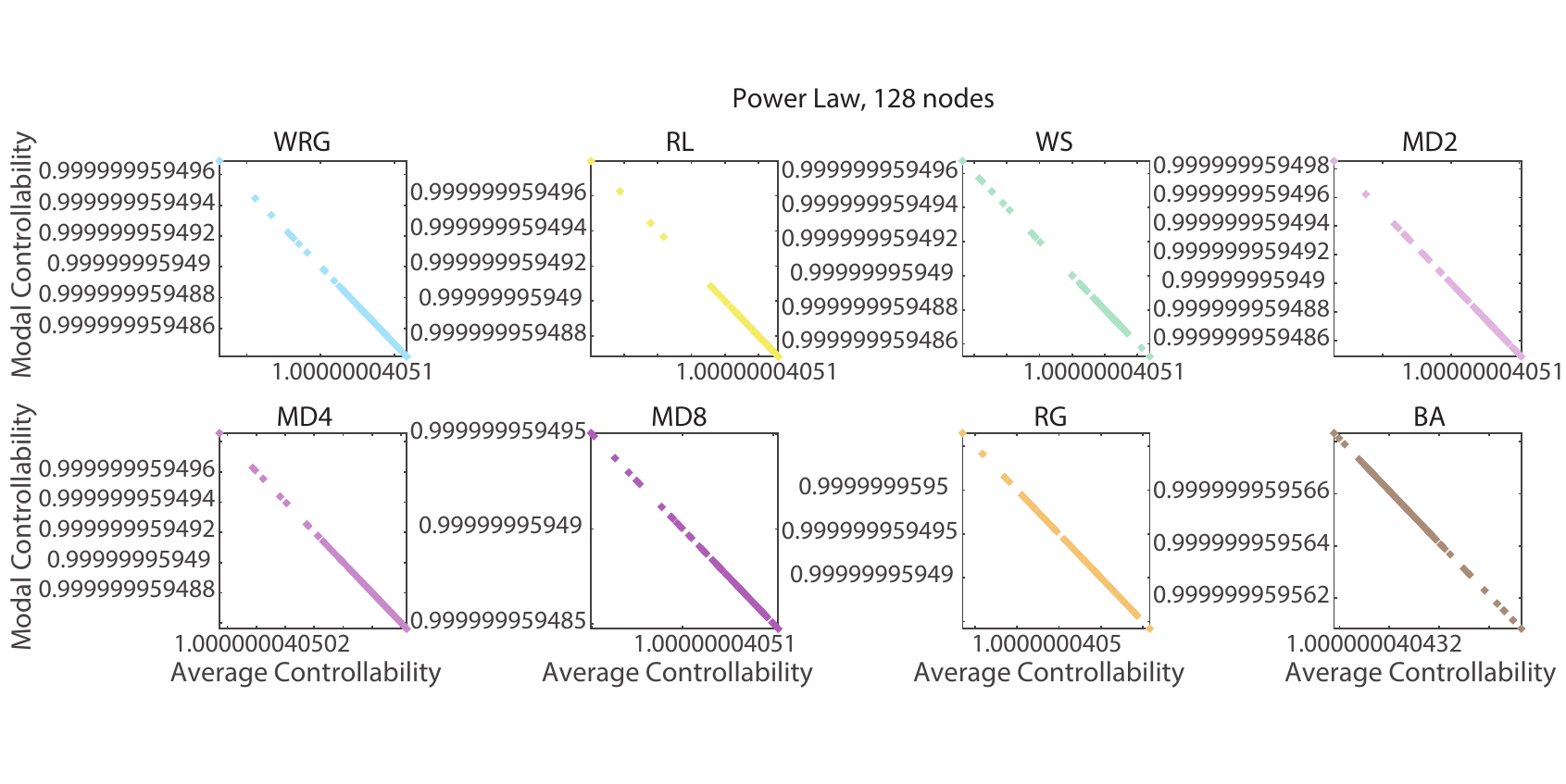}}
		\caption{\textbf{Relation between average and modal controllability across graphs for the power-law edge weighting scheme.} Average and modal controllability values were averaged over nodes in each graph model ensemble, and therefore scatterplots show values for each graph in the ensemble. Results are shown for the power-law edge weighting scheme. The eight graph models include the weighted random graph (WRG), the ring lattice (RL), the Watts-Strogatz small-world (WS), the modular graphs (MD2, MD4, MD8), the random geometric (RG), and the Barabasi-Albert preferential attachment (BA) models.  } \label{fig_P_mav128n}
		\vspace*{-10mm}
	\end{center}
\end{figure*}

\begin{figure*}
	\begin{center}
		\centerline{\includegraphics[width=0.9\textwidth]{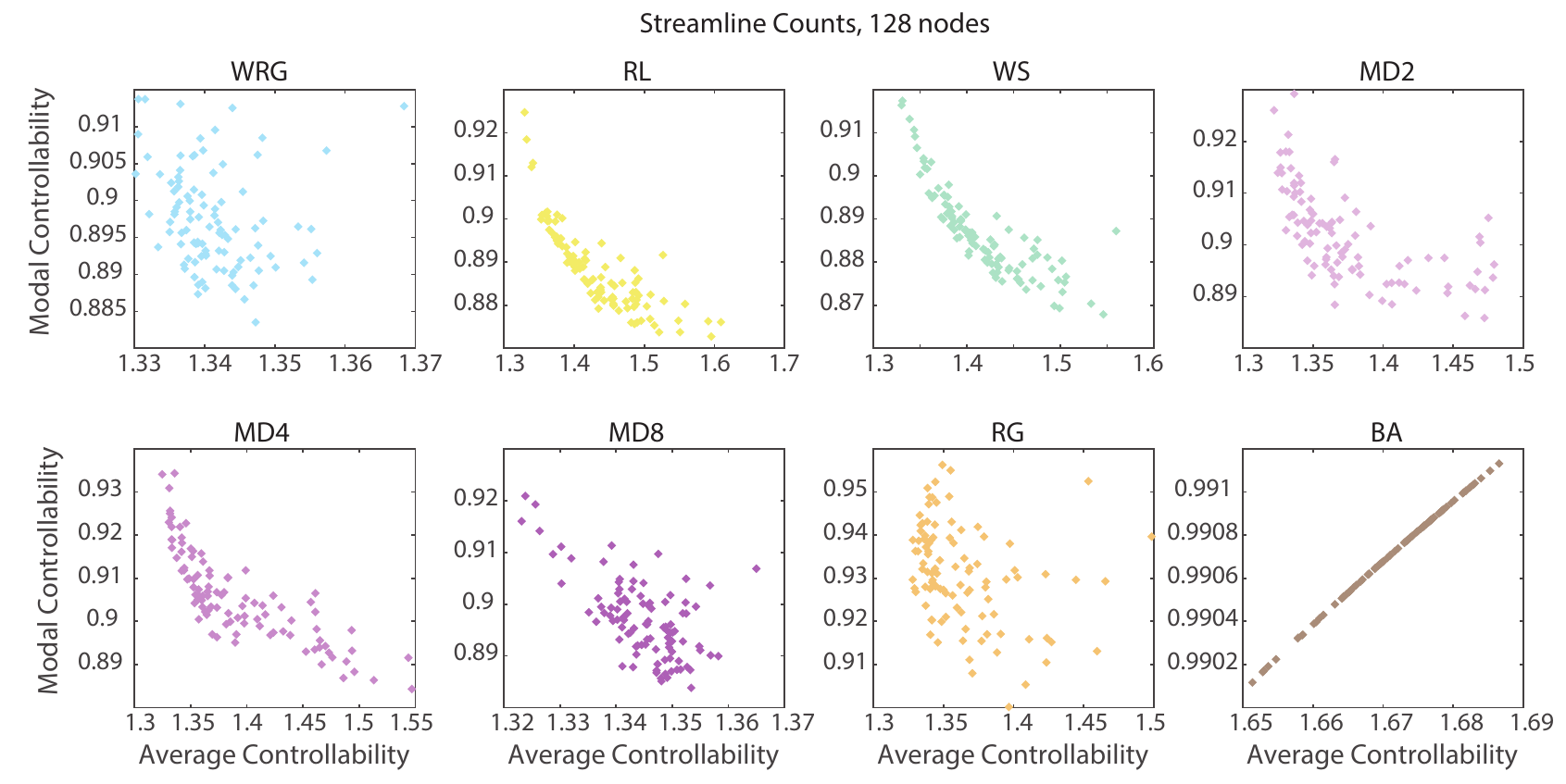}}
		\caption{\textbf{Relation between average and modal controllability across graphs for the streamline count edge weighting scheme.} Average and modal controllability values were averaged over nodes in each graph model ensemble, and therefore scatterplots show values for each graph in the ensemble. Results are shown for the streamline count edge weighting scheme. The eight graph models include the weighted random graph (WRG), the ring lattice (RL), the Watts-Strogatz small-world (WS), the modular graphs (MD2, MD4, MD8), the random geometric (RG), and the Barabasi-Albert preferential attachment (BA) models.  } \label{fig_str_mav128n}
		\vspace*{-10mm}
	\end{center}
\end{figure*}

\begin{figure*}
	\begin{center}
		\centerline{\includegraphics[width=0.9\textwidth]{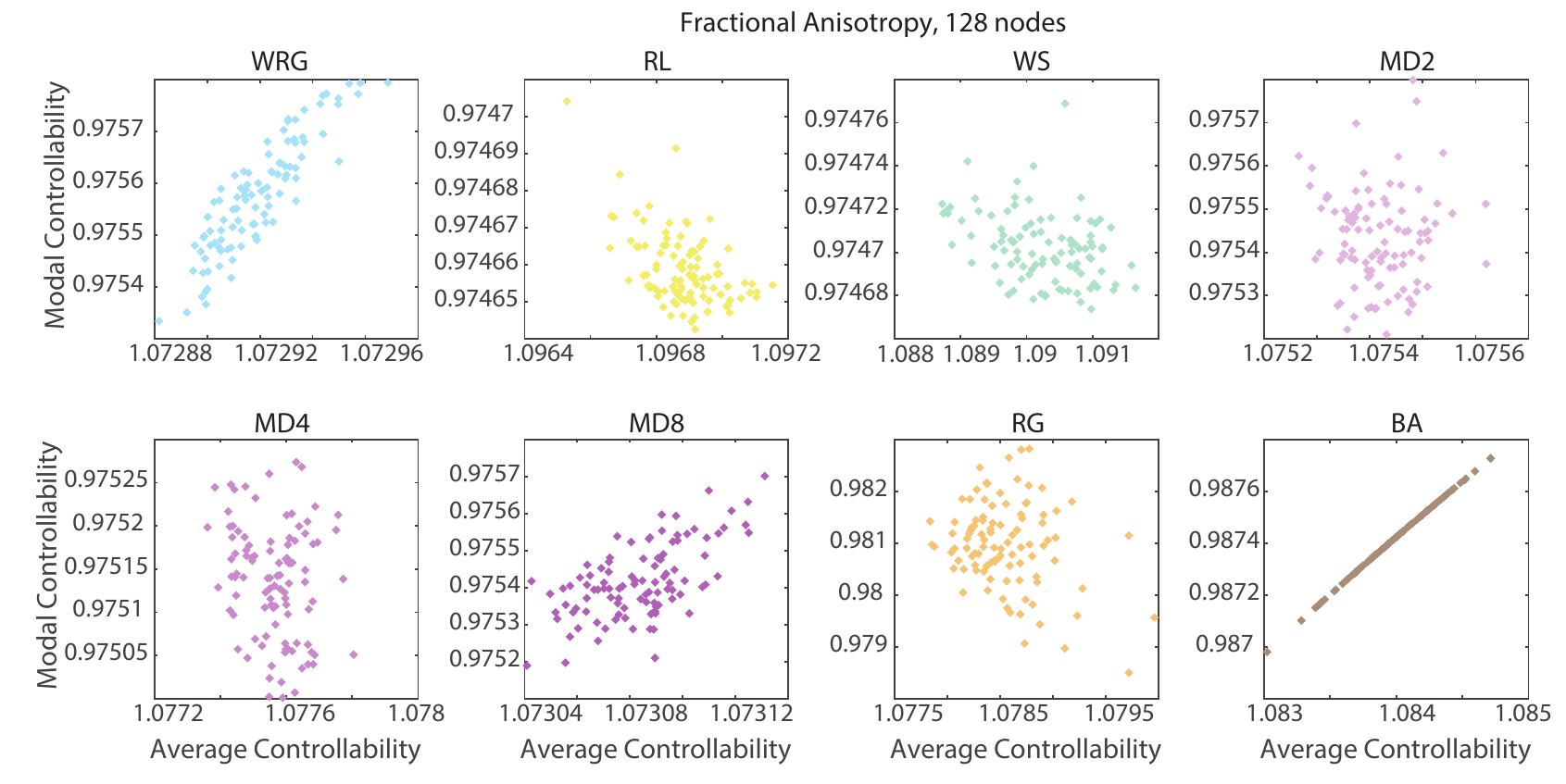}}
		\caption{\textbf{Relation between average and modal controllability across graphs for the FA edge weighting scheme.} Average and modal controllability values were averaged over nodes in each graph model ensemble, and therefore scatterplots show values for each graph in the ensemble. Results are shown for the FA edge weighting scheme. The eight graph models include the weighted random graph (WRG), the ring lattice (RL), the Watts-Strogatz small-world (WS), the modular graphs (MD2, MD4, MD8), the random geometric (RG), and the Barabasi-Albert preferential attachment (BA) models.  } \label{fig_FA_mav128n}
		\vspace*{-10mm}
	\end{center}
\end{figure*}

As noted in the main manuscript, in prior work we have observed that average and modal controllability, averaged over nodes, are positively related to one another across brain networks \cite{tang2016structural}.  These data suggest that brains that are structurally predisposed to be effective in moving network dynamics into easy-to-reach states (via average control) are also the brain that are structurally predisposed to be effective in moving network dynamics into difficult-to-reach states (via modal control). Here, we ask whether this positive relationship between average and modal controllability across networks holds in canonical graph models. In the main manuscript, we show scatter plots of average and modal controllability for graph models with Gaussian edge weight distributions. Here we show the complementary results for graph models with the power-law (Fig.~\ref{fig_P_mav128n}), streamline counts (Fig.~\ref{fig_str_mav128n}), and FA (Fig.~\ref{fig_FA_mav128n}) weighting schemes. Our results demonstrate that there is no consistent relationship between average and modal controllability across graph models and edge weight distributions; the relation between these two variables depends on both graph topology and graph geometry.

\subsubsection*{Relation between average and modal controllability across nodes}

\begin{figure*}
	\begin{center}
		\centerline{\includegraphics[width=0.9\textwidth]{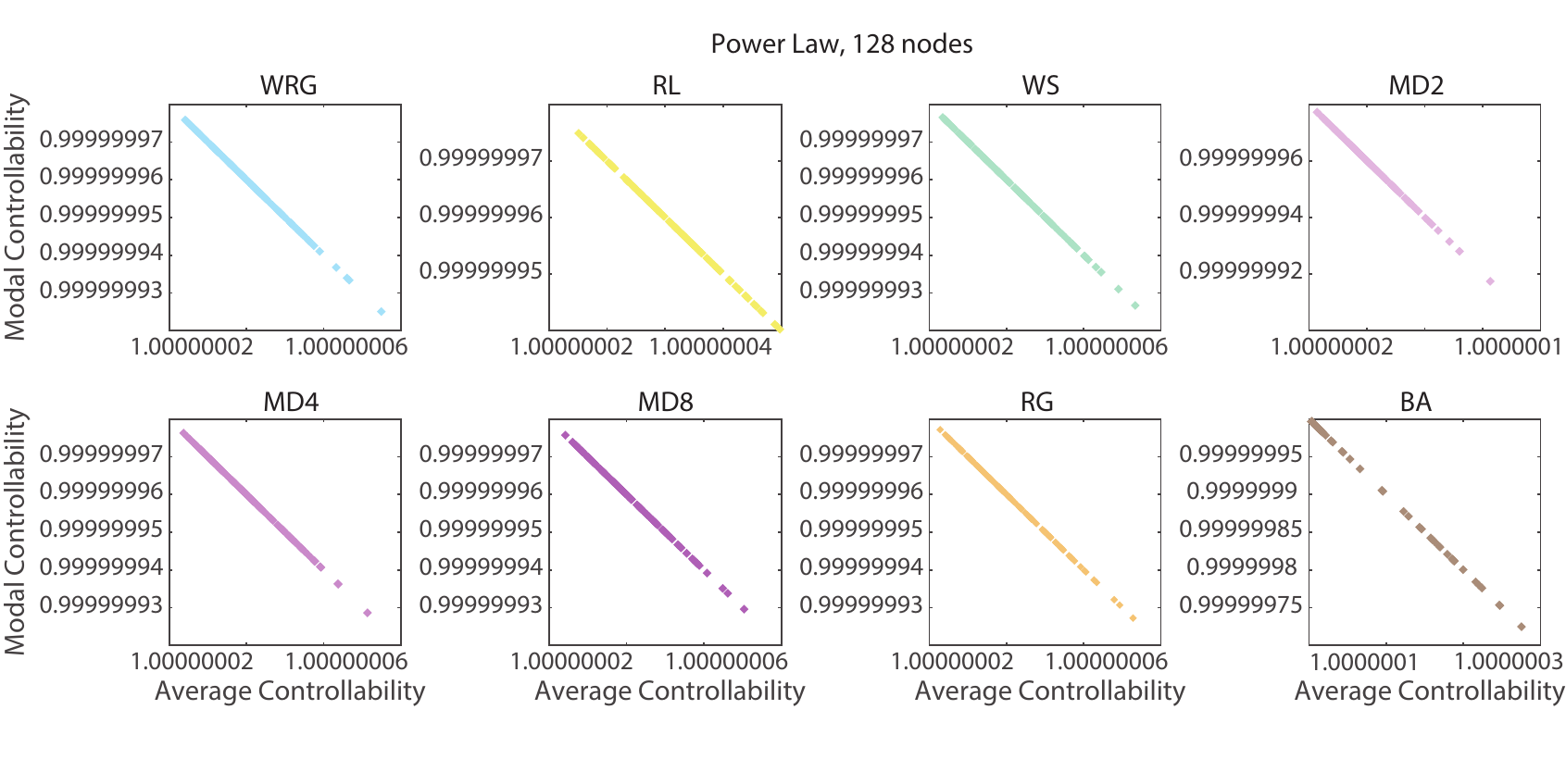}}
		\caption{\textbf{Relation between average and modal controllability across nodes for the power-law edge weighting scheme.} Average and modal controllability values were averaged over instances in each graph model ensemble, and therefore scatterplots show values for each node in the graph. Results are shown for the power-law edge weighting scheme. The eight graph models include the weighted random graph (WRG), the ring lattice (RL), the Watts-Strogatz small-world (WS), the modular graphs (MD2, MD4, MD8), the random geometric (RG), and the Barabasi-Albert preferential attachment (BA) models.  } \label{fig_P_mav128}
		\vspace*{-10mm}
	\end{center}
\end{figure*}

\begin{figure*}
	\begin{center}
		\centerline{\includegraphics[width=0.9\textwidth]{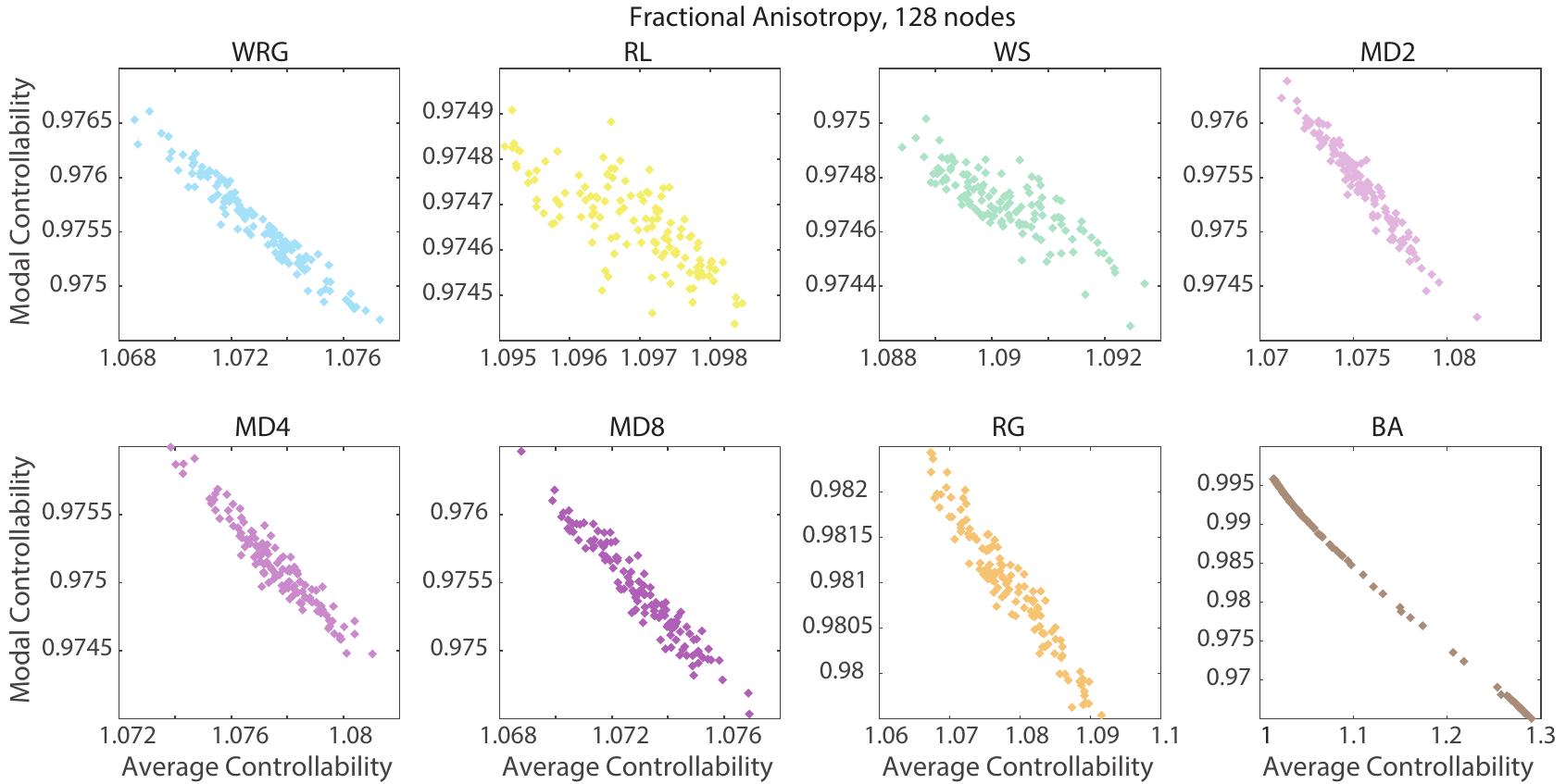}}
		\caption{\textbf{Relation between average and modal controllability across nodes for the FA edge weighting scheme.} Average and modal controllability values were averaged over instances in each graph model ensemble, and therefore scatterplots show values for each node in the graph. Results are shown for the FA edge weighting scheme. The eight graph models include the weighted random graph (WRG), the ring lattice (RL), the Watts-Strogatz small-world (WS), the modular graphs (MD2, MD4, MD8), the random geometric (RG), and the Barabasi-Albert preferential attachment (BA) models.  } \label{fig_FA_mav128}
		\vspace*{-10mm}
	\end{center}
\end{figure*}

\begin{figure*}
	\begin{center}
		\centerline{\includegraphics[width=0.9\textwidth]{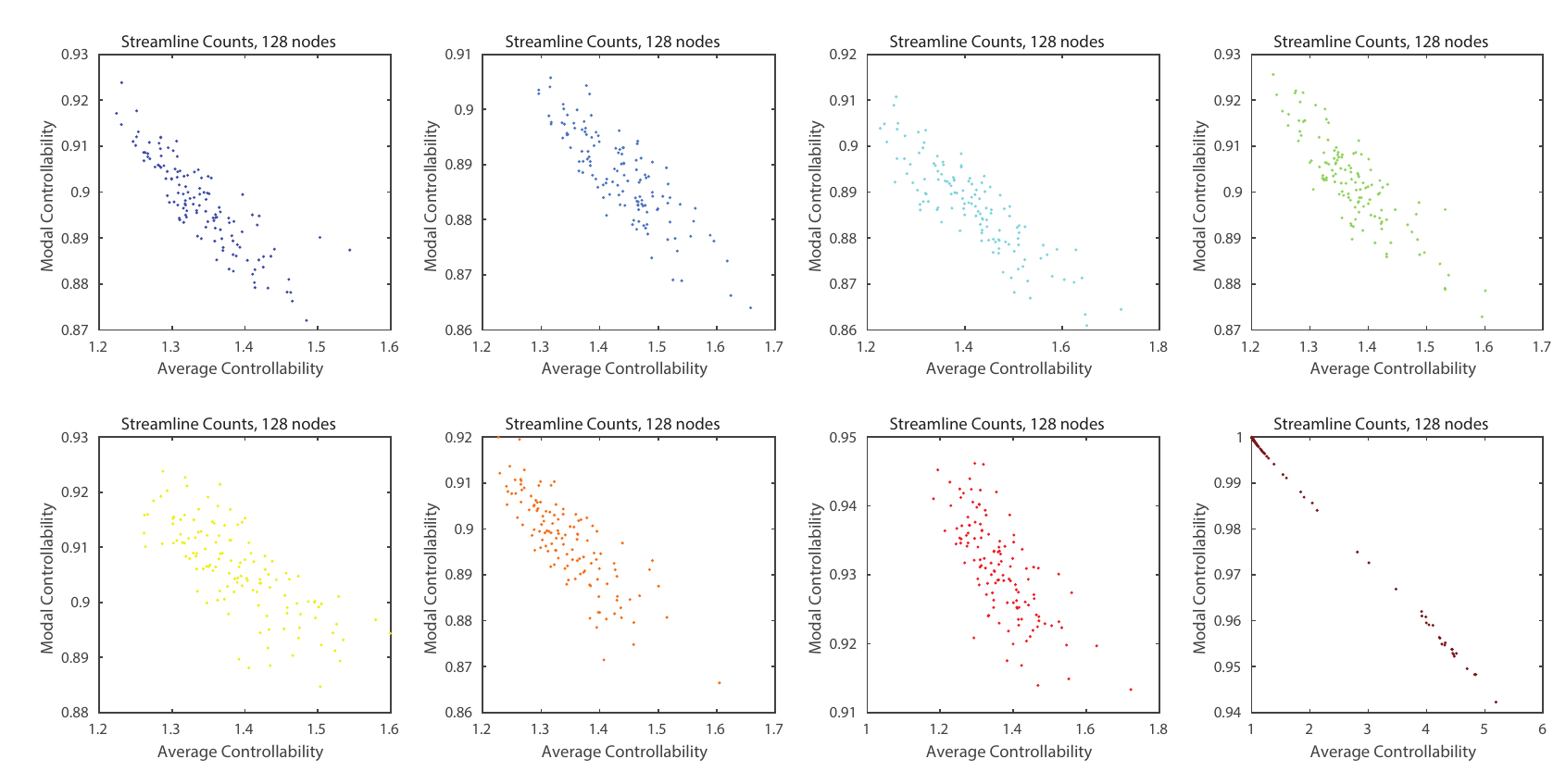}}
		\caption{\textbf{Relation between average and modal controllability across nodes for the streamline count edge weighting scheme.} Average and modal controllability values were averaged over instances in each graph model ensemble, and therefore scatterplots show values for each node in the graph. Results are shown for the streamline count edge weighting scheme. The eight graph models include the weighted random graph (WRG), the ring lattice (RL), the Watts-Strogatz small-world (WS), the modular graphs (MD2, MD4, MD8), the random geometric (RG), and the Barabasi-Albert preferential attachment (BA) models.  } \label{fig_str_mav128}
		\vspace*{-10mm}
	\end{center}
\end{figure*}

As noted in the main manuscript, in prior work we have observed that average and modal controllability are inversely related to one another across regions in brain networks, both in non-invasive human neuroimaging data acquired in youth and adults \cite{gu2015controllability,tang2016structural} and in tract-tracing data acquire from macaque monkeys \cite{gu2015controllability}.  These data suggest that regions of the brain that are structurally predisposed to be effective in moving the brain into easy-to-reach states (via average control) are different from the regions of the brain that are structural predisposed to be effective in moving the brain into difficult-to-reach states (via modal control). We ask the important more general question of whether this inverse relationship between average and modal controllability across nodes holds in canonical graph models.  In the main manuscript, we show scatter plots of average and modal controllability for graph models with Gaussian edge weight distributions. Here we show the complementary results for graph models with the power-law (Fig.~\ref{fig_P_mav128}), streamline counts (Fig.~\ref{fig_str_mav128}), and FA (Fig.~\ref{fig_FA_mav128}) weighting schemes. Our results demonstrate that the inverse relationship between average and modal controllability holds across graph models and edge weight distributions.

\subsection*{Effect of graph size on variation in network controllability statistics across graph models}

\begin{figure*}
	\begin{center}
		\centerline{\includegraphics[width=0.9\textwidth]{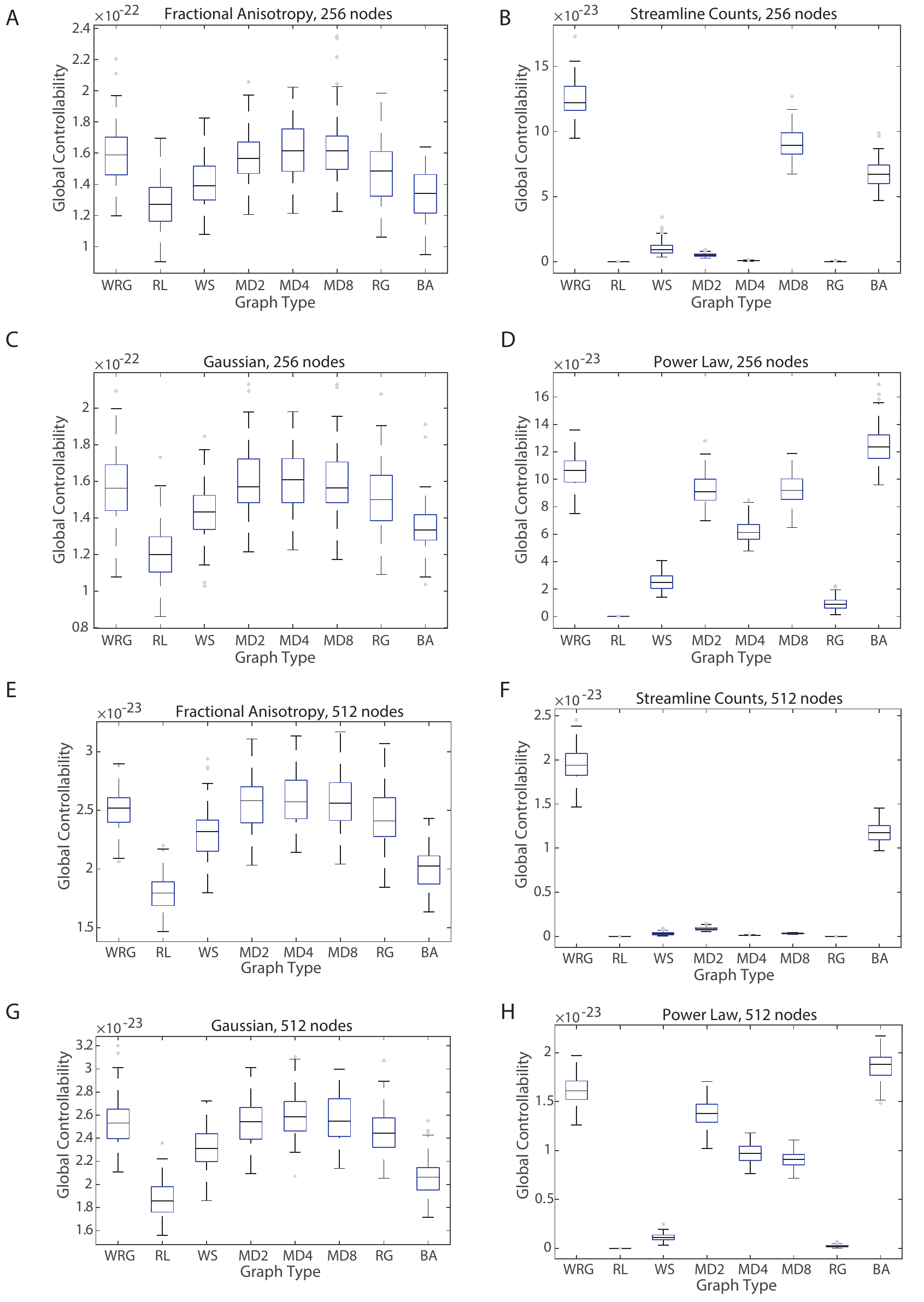}}
		\caption{\textbf{Variation in global controllability as a function of edge weighting and graph model.} Global controllability values were averaged over nodes in each graph model ensemble, and therefore boxplots show variation over graphs in the ensemble. Results are shown for four edge weighting schemes: \emph{(A,E)} FA, \emph{(B,F)} streamline counts, \emph{(C,G)} Gaussian, \emph{(D,H)} power-law. The eight graph models include the weighted random graph (WRG), the ring lattice (RL), the Watts-Strogatz small-world (WS), the modular graphs (MD2, MD4, MD8), the random geometric (RG), and the Barabasi-Albert preferential attachment (BA) models. Number of nodes is either 256 \emph{(A-D)} or 512 \emph{(E-H}). } \label{fig_g256n}
		\vspace*{-10mm}
	\end{center}
\end{figure*}

\begin{figure*}
	\begin{center}
		\centerline{\includegraphics[width=0.9\textwidth]{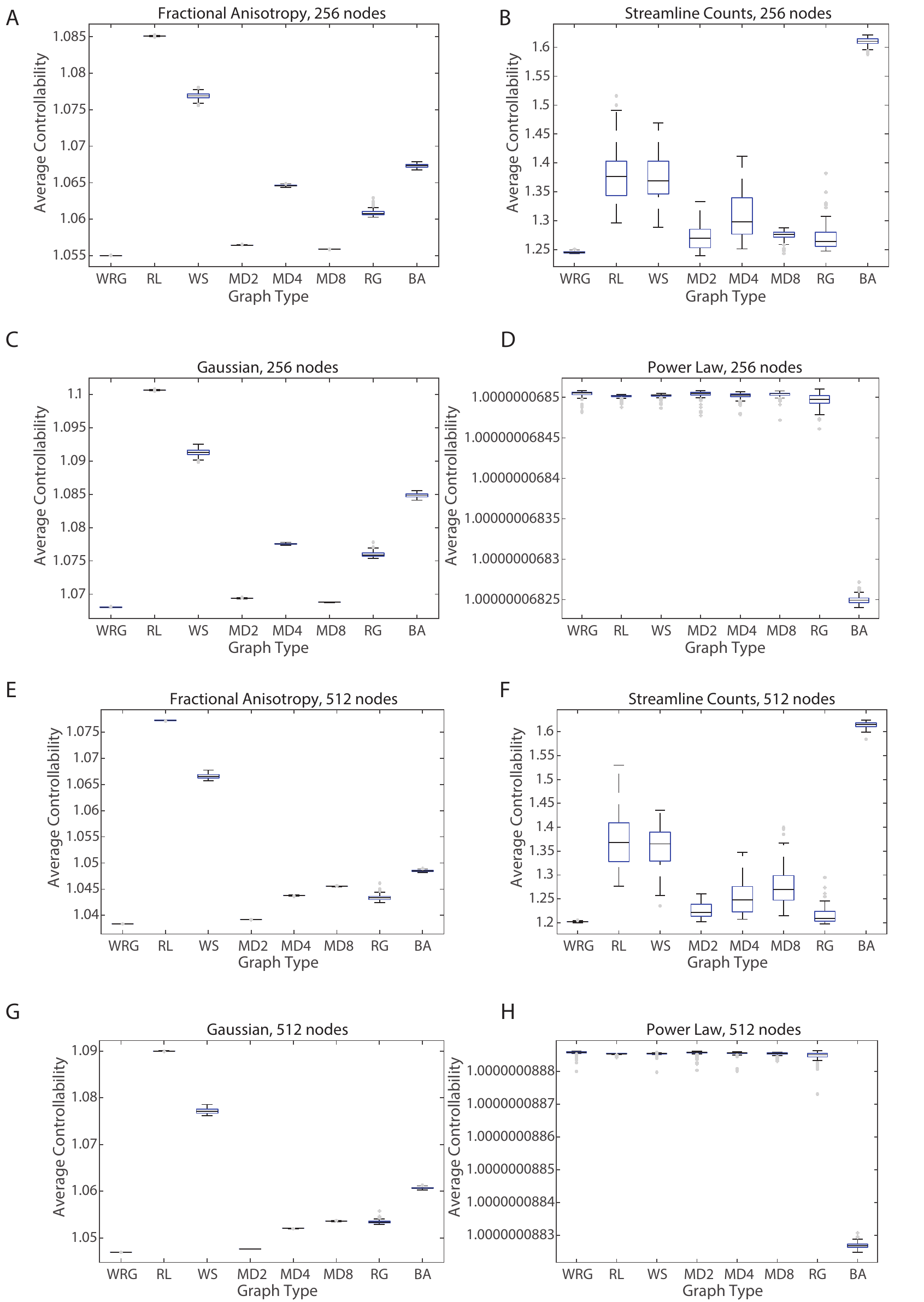}}
		\caption{\textbf{Variation in average controllability as a function of edge weighting and graph model.} Average controllability values were averaged over nodes in each graph model ensemble, and therefore boxplots show variation over graphs in the ensemble. Results are shown for four edge weighting schemes: \emph{(A,E)} FA, \emph{(B,F)} streamline counts, \emph{(C,G)} Gaussian, \emph{(D,H)} power-law. The eight graph models include the weighted random graph (WRG), the ring lattice (RL), the Watts-Strogatz small-world (WS), the modular graphs (MD2, MD4, MD8), the random geometric (RG), and the Barabasi-Albert preferential attachment (BA) models. Number of nodes is either 256 \emph{(A-D)} or 512 \emph{(E-H}). } \label{fig_a256n}
		\vspace*{-10mm}
	\end{center}
\end{figure*}

\begin{figure*}
	\begin{center}
		\centerline{\includegraphics[width=0.9\textwidth]{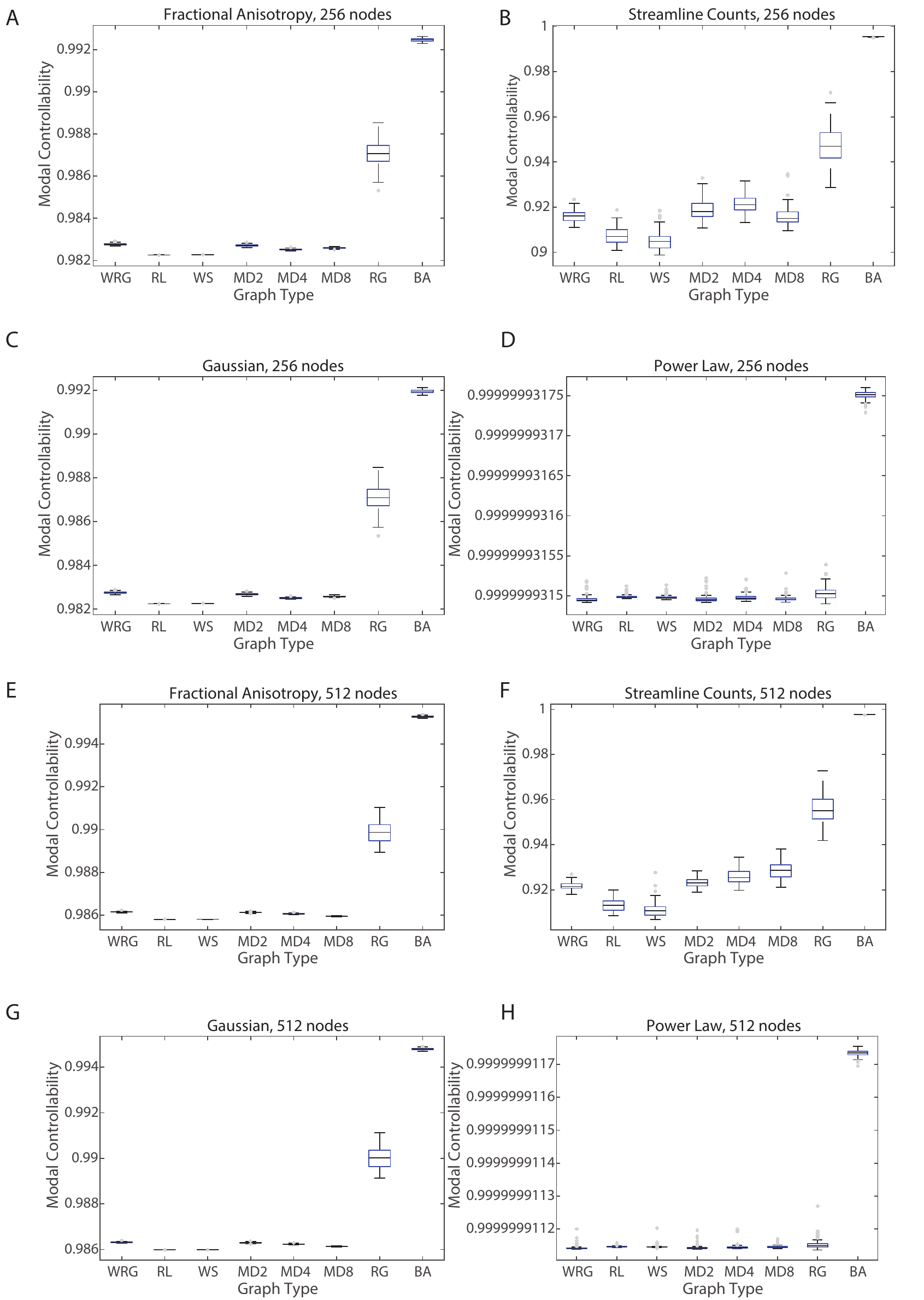}}
		\caption{\textbf{Variation in modal controllability as a function of edge weighting and graph model.} Modal controllability values were averaged over nodes in each graph model ensemble, and therefore boxplots show variation over graphs in the ensemble. Results are shown for four edge weighting schemes: \emph{(A,E)} FA, \emph{(B,F)} streamline counts, \emph{(C,G)} Gaussian, \emph{(D,H)} power-law. The eight graph models include the weighted random graph (WRG), the ring lattice (RL), the Watts-Strogatz small-world (WS), the modular graphs (MD2, MD4, MD8), the random geometric (RG), and the Barabasi-Albert preferential attachment (BA) models. Number of nodes is either 256 \emph{(A-D)} or 512 \emph{(E-H}). } \label{fig_m256n}
		\vspace*{-10mm}
	\end{center}
\end{figure*}

\begin{figure*}
	\begin{center}
		\centerline{\includegraphics[width=0.70\textwidth]{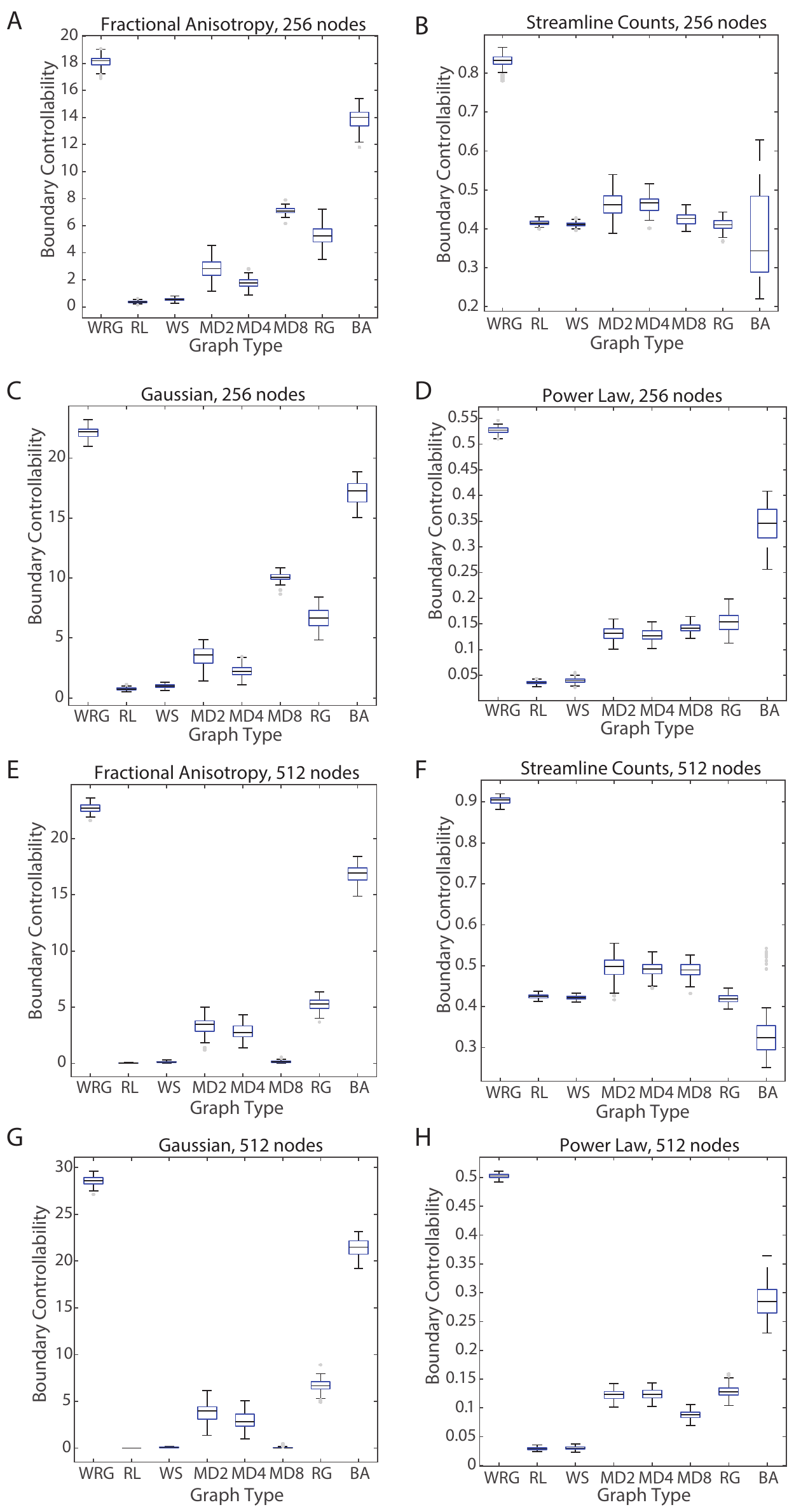}}
		\caption{\textbf{Variation in boundary controllability as a function of edge weighting and graph model.} Boundary controllability values were averaged over nodes in each graph model ensemble, and therefore boxplots show variation over graphs in the ensemble. Results are shown for four edge weighting schemes: \emph{(A,E)} FA, \emph{(B,F)} streamline counts, \emph{(C,G)} Gaussian, \emph{(D,H)} power-law. The eight graph models include the weighted random graph (WRG), the ring lattice (RL), the Watts-Strogatz small-world (WS), the modular graphs (MD2, MD4, MD8), the random geometric (RG), and the Barabasi-Albert preferential attachment (BA) models. Number of nodes is either 256 \emph{(A-D)} or 512 \emph{(E-H}). } \label{fig_b256n}
		\vspace*{-10mm}
	\end{center}
\end{figure*}

To assess the reliability and reproducibility of our results, we next examined the impact of graph size (n = 128, 256, or 512) on network controllability statistics, and their modulation by graph model for the Gaussian edge weight distribution. In this section, we examine how network controllability statistics vary over graphs within a given ensemble, and whether that variation differs as a function of graph model. Following the procedure outlined in the Methods section, for each controllability type, we took the mean of the $n$ sorted controllability values across the $n$ nodes in each graph instance, giving us 100 controllability values averaged over the $n$ regions. For each controllability type, we used four identical edge weight distributions corresponding to fractional anisotropy (FA) and streamline counts (SC) from real brain data, Gaussian distribution, and power law distribution. We observed that global, average, modal, and boundary controllability values for graphs of 256 and 512 nodes largely maintained the same trends as for graphs with 128 nodes (Fig.~\ref{fig_g256n}, ~\ref{fig_a256n}, ~\ref{fig_m256n}, and ~\ref{fig_b256n}).

\subsection*{Effect of graph size on nodal variation in network controllability statistics across graph models}

\begin{figure*}
	\begin{center}
		\centerline{\includegraphics[width=0.9\textwidth]{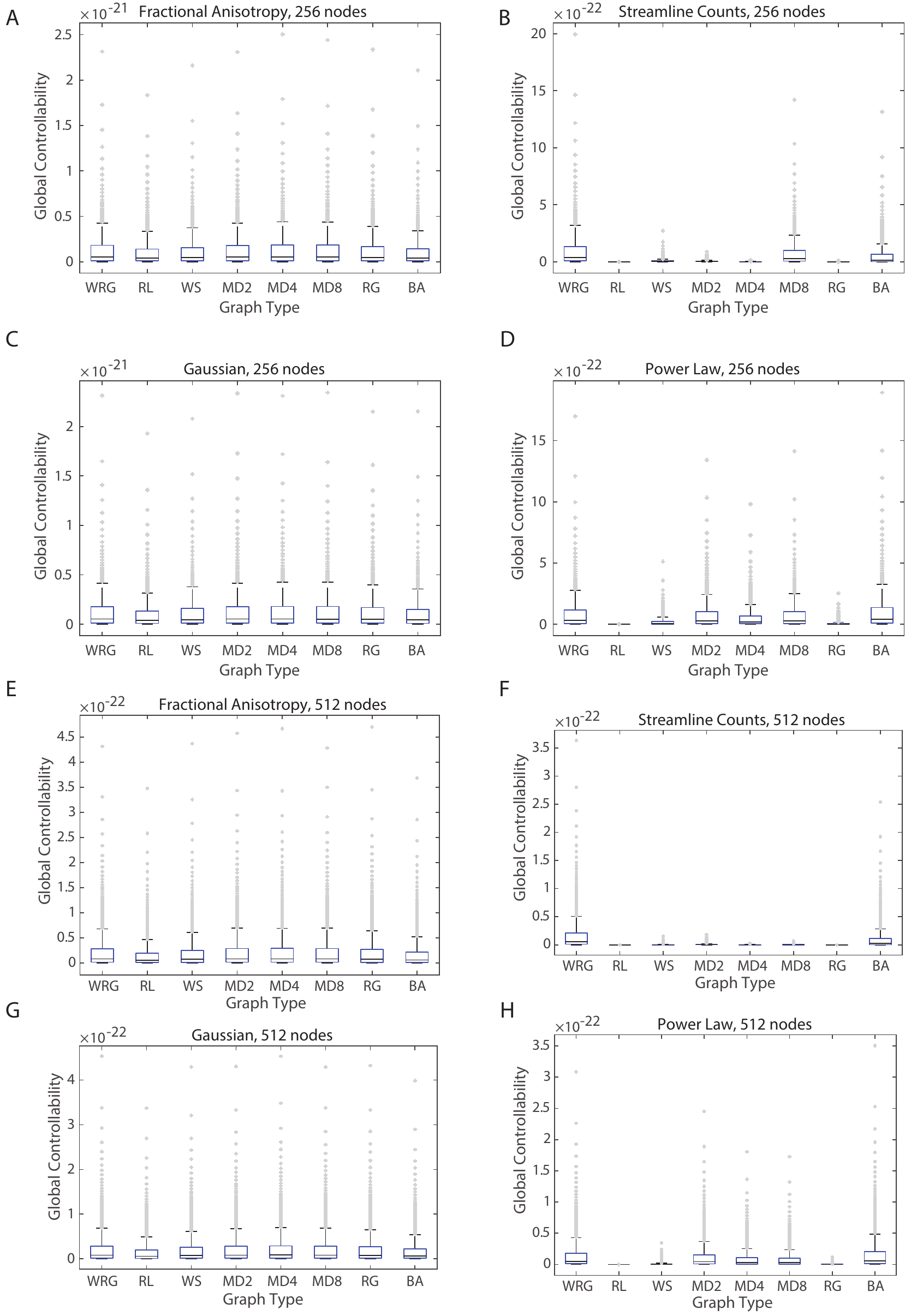}}
		\caption{\textbf{Nodal variation in global controllability as a function of edge weighting and graph model.} Global controllability values were averaged over instances in each graph model ensemble, and therefore boxplots show variation over nodes in the graph. Results are shown for four edge weighting schemes: \emph{(A,E)} FA, \emph{(B,F)} streamline counts, \emph{(C,G)} Gaussian, \emph{(D,H)} power-law. The eight graph models include the weighted random graph (WRG), the ring lattice (RL), the Watts-Strogatz small-world (WS), the modular graphs (MD2, MD4, MD8), the random geometric (RG), and the Barabasi-Albert preferential attachment (BA) models. Number of nodes is either 256 \emph{(A-D)} or 512 \emph{(E-H}). } \label{fig_g256}
		\vspace*{-10mm}
	\end{center}
\end{figure*}

\begin{figure*}
	\begin{center}
		\centerline{\includegraphics[width=0.9\textwidth]{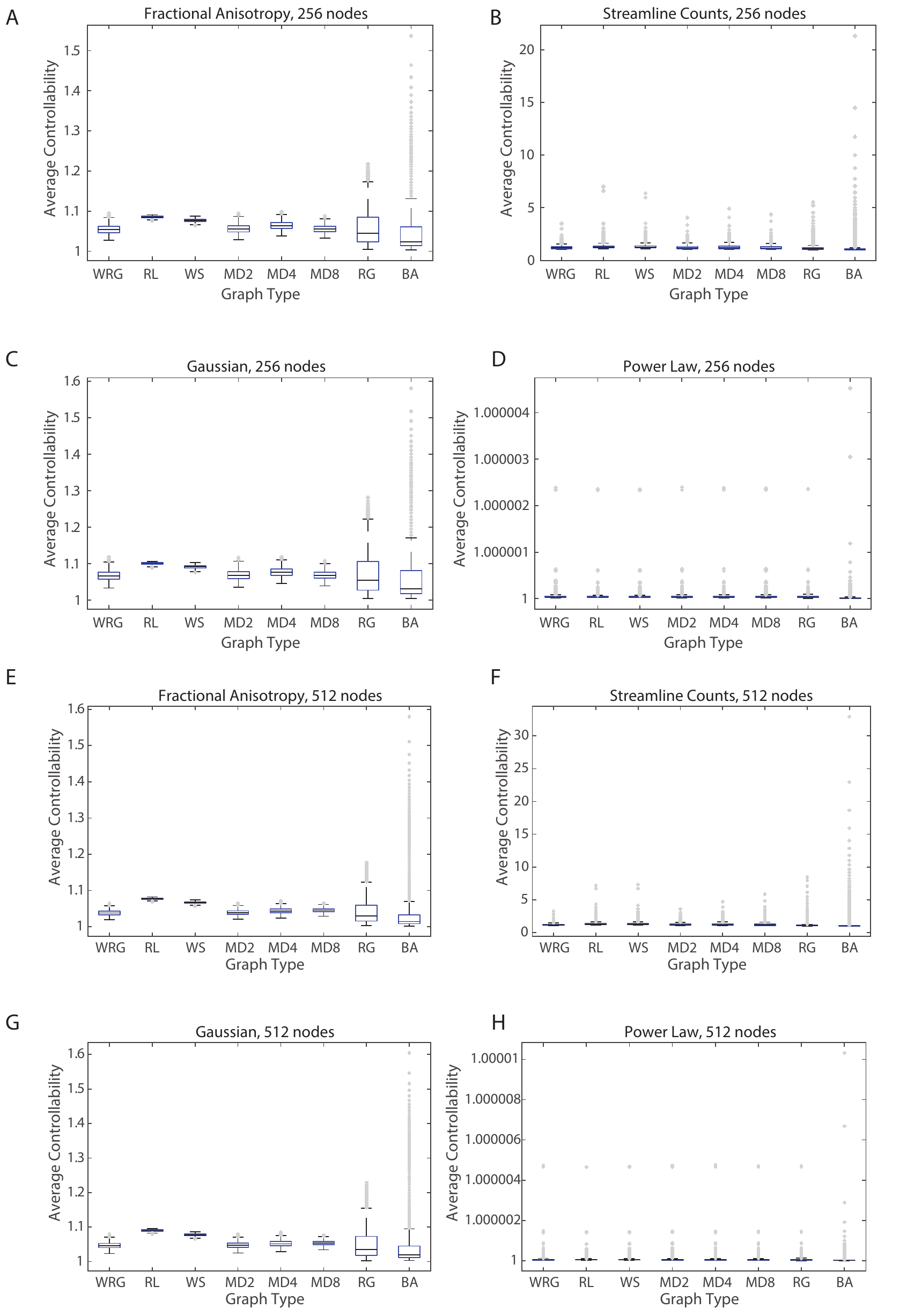}}
		\caption{\textbf{Nodal variation in average controllability as a function of edge weighting and graph model.} Average controllability values were averaged over instances in each graph model ensemble, and therefore boxplots show variation over nodes in the graph. Results are shown for four edge weighting schemes: \emph{(A,E)} FA, \emph{(B,F)} streamline counts, \emph{(C,G)} Gaussian, \emph{(D,H)} power-law. The eight graph models include the weighted random graph (WRG), the ring lattice (RL), the Watts-Strogatz small-world (WS), the modular graphs (MD2, MD4, MD8), the random geometric (RG), and the Barabasi-Albert preferential attachment (BA) models. Number of nodes is either 256 \emph{(A-D)} or 512 \emph{(E-H}). } \label{fig_a256}
		\vspace*{-10mm}
	\end{center}
\end{figure*}

\begin{figure*}
	\begin{center}
		\centerline{\includegraphics[width=0.9\textwidth]{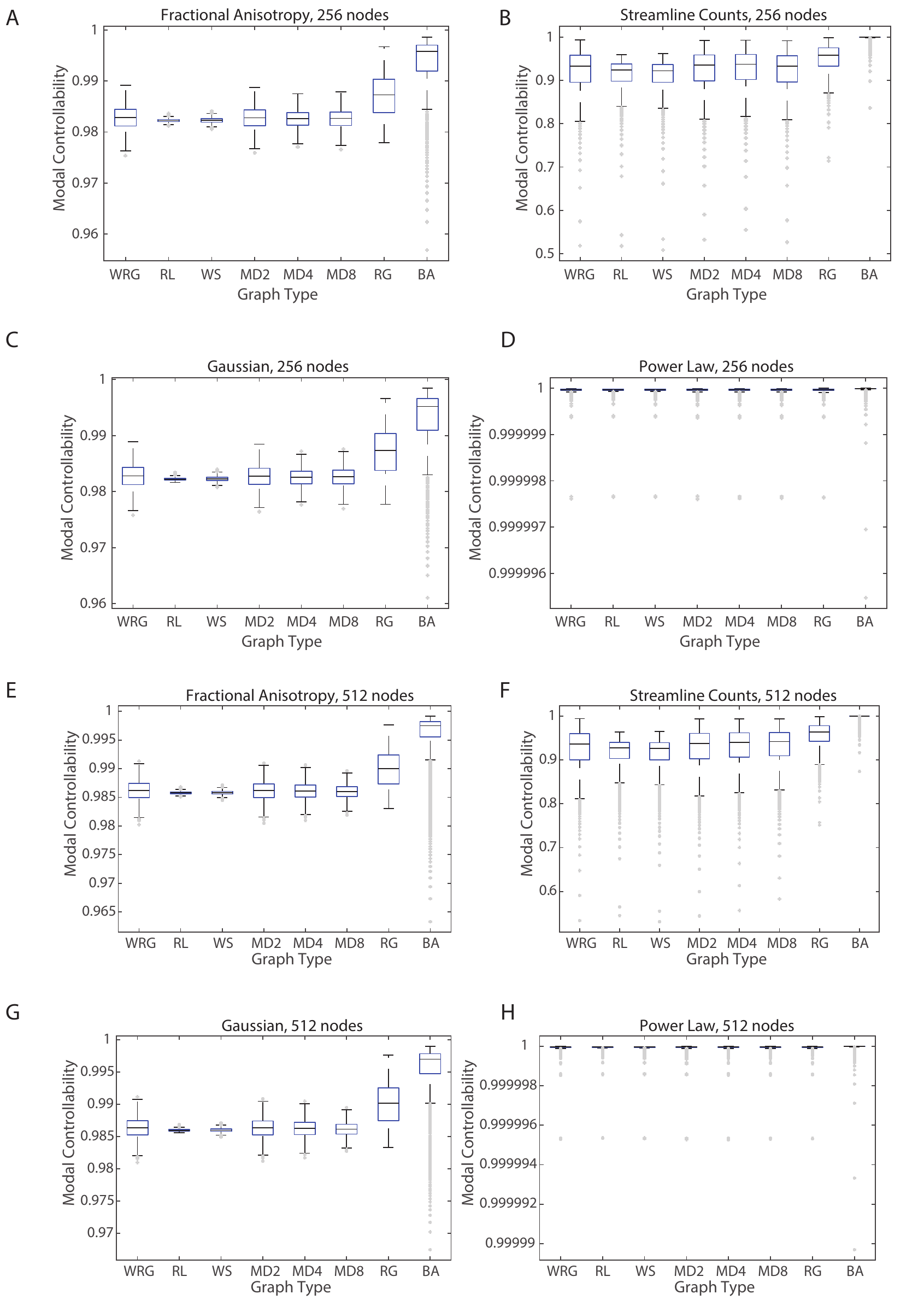}}
		\caption{\textbf{Nodal variation in modal controllability as a function of edge weighting and graph model.} Modal controllability values were averaged over instances in each graph model ensemble, and therefore boxplots show variation over nodes in the graph. Results are shown for four edge weighting schemes: \emph{(A,E)} FA, \emph{(B,F)} streamline counts, \emph{(C,G)} Gaussian, \emph{(D,H)} power-law. The eight graph models include the weighted random graph (WRG), the ring lattice (RL), the Watts-Strogatz small-world (WS), the modular graphs (MD2, MD4, MD8), the random geometric (RG), and the Barabasi-Albert preferential attachment (BA) models. Number of nodes is either 256 \emph{(A-D)} or 512 \emph{(E-H}). } \label{fig_m256}
		\vspace*{-10mm}
	\end{center}
\end{figure*}

\begin{figure*}
	\begin{center}
		\centerline{\includegraphics[width=0.70\textwidth]{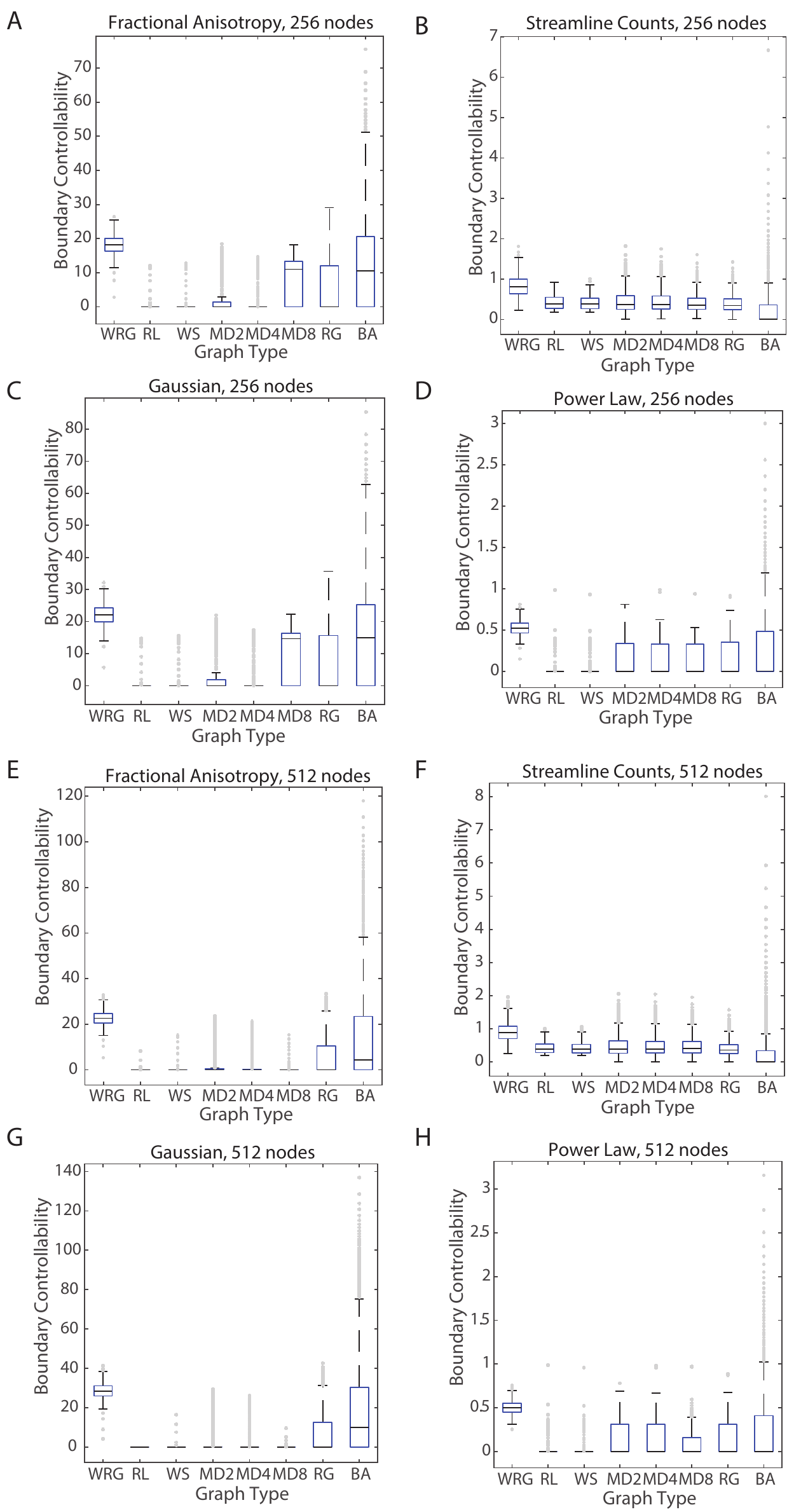}}
		\caption{\textbf{Nodal variation in boundary controllability as a function of edge weighting and graph model.} Boundary controllability values were averaged over instances in each graph model ensemble, and therefore boxplots show variation over nodes in the graph. Results are shown for four edge weighting schemes: \emph{(A,E)} FA, \emph{(B,F)} streamline counts, \emph{(C,G)} Gaussian, \emph{(D,H)} power-law. The eight graph models include the weighted random graph (WRG), the ring lattice (RL), the Watts-Strogatz small-world (WS), the modular graphs (MD2, MD4, MD8), the random geometric (RG), and the Barabasi-Albert preferential attachment (BA) models. Number of nodes is either 256 \emph{(A-D)} or 512 \emph{(E-H}). } \label{fig_b256}
		\vspace*{-10mm}
	\end{center}
\end{figure*}

Here we again assess the reliability and reproducibility of our results, by examining the impact of graph size (n = 128, 256, or 512) on network controllability statistics, and their modulation by graph model for the Gaussian edge weight distribution. In this section, we examine the nodal variation in network controllability statistics, and ask questions regarding how nodal variation differs across graph models, and between controllability statistics. Following the procedure outlined in the Methods section, for each controllability type we took the mean of the $n$ sorted nodal controllability values across all 100 graphs in a given model ensemble, giving us $n$ controllability values averaged over the graph instances. For each controllability type, we used four identical edge weight distributions corresponding to fractional anisotropy (FA) and streamline counts (SC) from real brain data, Gaussian distribution, and power law distribution. We observed that global, average, modal, and boundary controllability values for graphs of 256 and 512 nodes largely maintained the same trends as for graphs with 128 nodes (Fig.~\ref{fig_g256}, ~\ref{fig_a256}, ~\ref{fig_m256}, and ~\ref{fig_b256}).

\subsection*{Effect of $\rho$ on boundary controllability estimates}

\begin{figure*}
	\begin{center}
		\centerline{\includegraphics[width=0.9\textwidth]{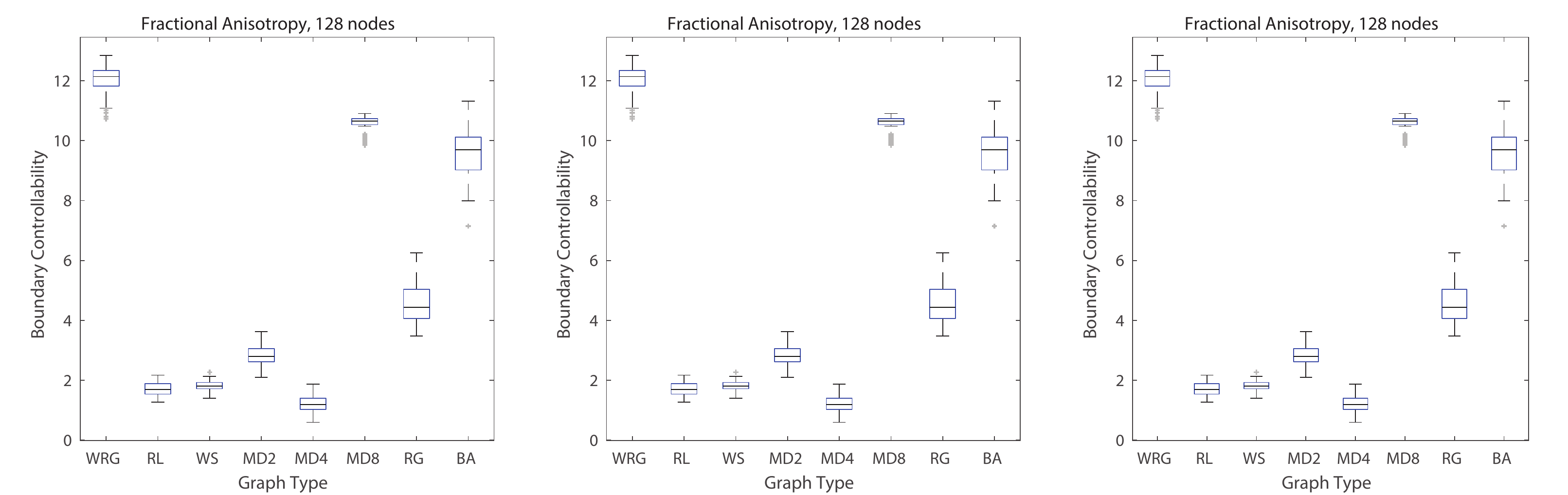}}
		\caption{\textbf{Robustness of boundary controllability estimates to variations in $\rho$.} \emph{(Left)} Estimates for variation in boundary controllability over graph instances for all eight graph models weighted using the Fractional Anisotropy edge weight distribution, and calculated using $\rho = 1 \times 10^{-3}$.  \emph{(Middle)} Estimates for variation in boundary controllability over graph instances for all eight graph models weighted using the Fractional Anisotropy edge weight distribution, and calculated using $\rho = 1 \times 10^{-5}$. \emph{(Right)} Estimates for variation in boundary controllability over graph instances for all eight graph models weighted using the Fractional Anisotropy edge weight distribution, and calculated using $\rho = 1 \times 10^{-8}$.} \label{fig_b128_rho}
		\vspace*{-10mm}
	\end{center}
\end{figure*}

In the main text, we provided estimates of boundary controllability using $\rho = 1 \times 10^{-5}$. However, we observe similar estimates of boundary controllability for $\rho = 1 \times 10^{-3}$ and $\rho = 1 \times 10^{-8}$. Here we provide an illustrative example for the 128-node graphs using the Fractional Anisotropy edge weight distribution (Fig.~\ref{fig_b128_rho}).

\clearpage
\newpage

\bibliography{../bibfile}


\end{document}